\documentclass[12pt,preprint]{aastex}

\newcommand{\be}{\begin{equation}}
\newcommand{\ee}{\end{equation}}
\newcommand{\ul}{\underline{\hspace{40pt}}}
\newcommand{\bea}{\begin{eqnarray}}
\newcommand{\eea}{\end{eqnarray}}
\newcommand{\tbnm}{\tablenotemark}
\newcommand{\tbnt}{\tablenotetext}

\newcommand{\cmMMM}{\mbox{cm$^{-3}$}}
\newcommand{\gcmMMM}{\mbox{g\,cm$^{-3}$}}
\newcommand{\cc}{\mbox{cm$^{-3}$}}
\newcommand{\kms}{\mbox{km s$^{-1}$}}
\newcommand{\cm}{\mbox{cm}}
\newcommand{\yr}{\mbox{yr}}

\newcommand{\simlt}{\lesssim}
\newcommand{\simgt}{\gtrsim}

\newcommand{\real}{\mbox{\rm{Re}}}
\newcommand{\imag}{\mbox{\rm{Im}}}

\newcommand{\xhat}{\mbox{\boldmath{$\hat{x}$}}}

\newcommand{\kvec}{\mbox{\boldmath{$k$}}}

\newcommand{\Bvec}{\mbox{\boldmath{$B$}}}

\newcommand{\kB}{\mbox{$k_{\rm B}$}}

\newcommand{\vs}{\mbox{$v_{\rm{s}}$}}
\newcommand{\vsi}{\mbox{$v_{\rm{s},i}$}}

\newcommand{\Mga}{\mbox{$M_{\rm gA}$}}
\newcommand{\Mna}{\mbox{$M_{\rm nA}$}}

\newcommand{\cri}{\mbox{$\zeta_{{}_{\rm{CR}}}$}}

\newcommand{\vcrit}{\mbox{$v_{\rm crit}$}}
\newcommand{\vsig}{\mbox{$v_{\rm sig}$}}
\newcommand{\ellb}{\mbox{$\ell_B$}}

\newcommand{\nH}{\mbox{$n_{\rm H}$}}

\newcommand{\Htwo}{\mbox{H$_2$}}

\newcommand{\tdragij}{\mbox{$\tau^{\rm (i,j)}_{{\rm drag}}$}}
\newcommand{\tdragitot}{\mbox{$\tau^{\rm (i,tot)}_{{\rm drag}}$}}

\newcommand{\rhoc}{\mbox{$\rho_{\rm c}$}}

\newcommand{\nn}{\mbox{$n_{\rm n}$}}

\newcommand{\rhon}{\mbox{$\rho_{\rm n}$}}

\newcommand{\tdragni}{\mbox{$\tau^{\rm (n,i)}_{{\rm drag}}$}}

\newcommand{\tdragntot}{\mbox{$\tau^{\rm (n,tot)}_{{\rm drag}}$}}
\newcommand{\tdragng}{\mbox{$\tau^{\rm (n,g)}_{{\rm drag}}$}}

\newcommand{\mi}{\mbox{$m_{\rm i}$}}
\newcommand{\nni}{\mbox{$n_{\rm i}$}}

\newcommand{\rhoi}{\mbox{$\rho_{\rm i}$}}

\newcommand{\aplus}{\mbox{a$^+$}}
\newcommand{\maplus}{\mbox{$m_{{\rm a}^+}$}}
\newcommand{\mplus}{\mbox{m$^+$}}
\newcommand{\mmplus}{\mbox{$m_{{\rm m}^+}$}}

\newcommand{\vi}{\mbox{$v_{{\rm i}}$}}

\newcommand{\Omegi}{\mbox{$\Omega_{\rm i}$}}
\newcommand{\tdragin}{\mbox{$\tau^{\rm (i,n)}_{{\rm drag}}$}}

\newcommand{\nne}{\mbox{$n_{\rm e}$}}

\newcommand{\ag}{\mbox{$a_{\rm g}$}}
\newcommand{\agfr}{\mbox{$a_{\rm g,fr}$}}

\newcommand{\rhog}{\mbox{$\rho_{\rm g}$}}

\newcommand{\Omegag}{\mbox{$\Omega_{\rm g}$}}
\newcommand{\tdraggn}{\mbox{$\tau^{\rm (g,n)}_{{\rm drag}}$}}

\newcommand{\gminus}{\mbox{g$^-$}}
\newcommand{\Hallg}{\mbox{$\Gamma_{\rm g}$}}

\newcommand{\ngm}{\mbox{$n_{{\rm g}^-}$}}
\newcommand{\xgm}{\mbox{$x_{{\rm g}^-}$}}

\newcommand{\vg}{\mbox{$v_{\rm g}$}}

\newcommand{\gzero}{\mbox{g$^{0}$}}
\newcommand{\xgz}{\mbox{$x_{{\rm g}^0}$}}

\newcommand{\tdraggzn}{\mbox{$\tau^{\rm (g^{0},n)}_{{\rm drag}}$}}

\newcommand{\gplus}{\mbox{g$^{+}$}}
\newcommand{\ngp}{\mbox{$n_{{\rm g}^+}$}}
\newcommand{\xgp}{\mbox{$x_{{\rm g}^+}$}}

\newcommand{\sgminus}{\mbox{sg$^{-}$}}

\newcommand{\sgzero}{\mbox{sg$^{0}$}}

\newcommand{\sgplus}{\mbox{sg$^{+}$}}

\newcommand{\vphi}{\mbox{$v_{\phi}$}}
\newcommand{\omegar}{\mbox{$\omega_{\rm r}$}}
\newcommand{\omegai}{\mbox{$\omega_{\rm i}$}}
\newcommand{\tdamp}{\mbox{$\tau_{\rm damp}$}}

\newcommand{\vims}{\mbox{$V_{\rm ims}$}}
\newcommand{\taudims}{\mbox{$\tau_{\rm damp}^{\rm (ims)}$}}

\newcommand{\lambdaims}{\mbox{$\lambda_{\rm max}^{\rm (ims)}$}}

\newcommand{\Qeff}{\mbox{$Q_{\rm eff}$}}
\newcommand{\Omegaqp}{\mbox{$\Omega_{\rm qp}$}}

\newcommand{\lambdael}{\mbox{$\lambda_{\rm elect}$}}

\newcommand{\vgms}{\mbox{$V_{\rm gms}$}}
\newcommand{\vgmms}{\mbox{$V_{{\rm g^{-}ms}}$}}
\newcommand{\taudgms}{\mbox{$\tau_{\rm damp}^{\rm (gms)}$}}
\newcommand{\lambdamingms}{\mbox{$\lambda_{\rm min}^{\rm (gms)}$}}
\newcommand{\lambdamaxgms}{\mbox{$\lambda_{\rm max}^{\rm (gms)}$}}

\newcommand{\vnas}{\mbox{$C_{\rm ad}$}}
\newcommand{\vnis}{\mbox{$C_{\rm n}$}}
\newcommand{\Ciso}{\mbox{$C_{\rm{n}}$}}
\newcommand{\taudns}{\mbox{$\tau_{\rm damp}^{\rm (ns)}$}}

\newcommand{\vnms}{\mbox{$V_{\rm nms}$}}
\newcommand{\taudnms}{\mbox{$\tau_{\rm damp}^{\rm (nms)}$}}
\newcommand{\lambdaminnms}{\mbox{$\lambda_{\rm min}^{\rm (nms)}$}}

\newcommand{\Alf}{\mbox{Alfv\'{e}n}}

\newcommand{\vnA}{\mbox{$V_{\rm{nA}}$}}
\newcommand{\vgA}{\mbox{$V_{\rm gA}$}}
\newcommand{\viA}{\mbox{$V_{\rm{iA}}$}}

\newcommand{\Bxo}{\mbox{$B_{x,0}$}}
\newcommand{\rhoio}{\mbox{$\rho_{\rm i, 0}$}}

\shorttitle{Dust and the Critical Speed for C Shocks}
\shortauthors{G. E. Ciolek, W. G. Roberge, \& T. Ch. Mouschovias}
\slugcomment{Accepted by {\sl The Astrophysical Journal}}

\received{2002 July 27}
\begin{document}

\title{Multifluid, Magnetohydrodynamic Shock Waves with
Grain Dynamics II.  Dust and the
Critical Speed for C Shocks}

\author{Glenn E. Ciolek,\altaffilmark{1,2}
Wayne G. Roberge,\altaffilmark{1,2} and
Telemachos Ch. Mouschovias\altaffilmark{3}}

\altaffiltext{1}{New York Center for Studies on the Origin of Life (NSCORT)}
\altaffiltext{2}{Department of Physics, Applied Physics and Astronomy,
Rensselaer Polytechnic Institute,  110 8th Street, Troy, NY 12180}
\altaffiltext{3}{Departments of Physics and Astronomy, University of Illinois
at Urbana-Champaign, 1002 West Green Street, Urbana, IL 61801}
\email{cioleg@rpi.edu, roberw@rpi.edu}

\begin{abstract}
This is the second in a series of papers on the effects of dust on 
multifluid, magnetohydrodynamic shock waves in weakly-ionized molecular
gas. We investigate the influence of dust on the critical shock speed,
\vcrit, above which C shocks cease to exist.
Chernoff showed that \vcrit\ cannot exceed the grain magnetosound speed,
\vgms, if dust grains are dynamically well coupled to
the magnetic field. Since $\vgms \simeq 5$\,\kms\ in a dense cloud or
core, the potential implications for models of shock emission are 
profound. We present numerical simulations of steady shocks
where the grains may be well- or poorly coupled to the field.
We use a time-dependent, multifluid MHD code that models the plasma as a
system of interacting fluids: neutral particles, ions, electrons, and 
various ``dust fluids'' comprised of grains with different sizes and 
charges. Our simulations include grain inertia and grain charge 
fluctuations but to highlight the essential physics we assume adiabatic
flow, single-size grains, and neglect the effects of chemistry. We show
that the existence of a phase speed $\vphi$ does not necessarily mean
that C shocks will form for all shock speeds $\vs$ less than 
$\vphi$. When the grains are weakly coupled to the field, 
steady, adiabatic shocks resemble shocks with no dust: the transition to
J~type flow occurs at $\vcrit\approx 2.76\vnA$, where \vnA\ is the 
neutral \Alf\ speed, and steady shocks with $\vs>2.76\vnA$ are 
J~shocks with magnetic precursors in the ion-electron fluid. When the 
grains are strongly coupled to the field,
$\vcrit=\min\left(2.76\vnA,\vgms\right)$. Shocks with $\vcrit<\vs<\vgms$
have magnetic precursors in the ion-electron-dust fluid. Shocks with
$\vs>\vgms$ have no magnetic precursor in any fluid. We present 
time-dependent calculations to study the formation of steady multifluid
shocks. The dynamics differ qualitatively depending on whether or not
the grains and field are well coupled.
\end{abstract}



\section{Introduction}

Shock waves in weakly-ionized plasmas, such as the molecular outflows
around young stellar objects, may have a multifluid structure,
in which the
charged and neutral components of the plasma behave as distinct fluids.
If the shock speed is less than the speed with which the charged fluid
communicates compressive disturbances, the charged particles are 
accelerated ahead of the neutral gas in a ``magnetic precursor''
(Mullan 1971; Draine 1980). The streaming of charged particles through 
the neutrals also accelerates, compresses and heats the neutral fluid. 
If it remains sufficiently cool, the neutral
fluid is everywhere supersonic in a frame at rest in the post-shock gas.
In this case all hydrodynamical variables are continuous and the
entire structure is termed a ``C~shock'' (Draine 1980).
Alternatively the neutral fluid must undergo a transonic transition.
This may occur smoothly via continous flow through a sonic point
(a ``C$^*$ shock'', see Chernoff 1987 and Roberge \& Draine 1990, but
note that Chi\`{e}ze, Pineau des For\^{e}ts, \& Flower 1998
question the reality of C$^*$ shocks) or abruptly
at a viscous jump front (a ``J shock'').

Simulations of the different types of multifluid MHD shocks
reveal distinct observational signatures, including the strengths
and velocity profiles of atomic and molecular emission lines
(Chernoff, Hollenbach, \& McKee 1982; Draine \& Roberge 1982;
Draine, Roberge, \& Dalgarno 1983, henceforth DRD83; Smith \& Brand
1990a,b; Smith 1991a,b; Smith, Brand, \& Moorhouse 1991a,b; Kaufman
\& Neufeld 1996a,b; Timmermann 1996; Neufeld \& Stone 1997; Pineau des
For\^{e}ts et al.\ 2001), the ortho/para ratio of H$_2$ (Timmermann 
1998, Wilgenbus et al.\ 2000), and the abundances of various chemical
species in the post-shock gas (e.g., DRD83; Flower, Pineau des 
For\^{e}ts, \& Hartquist 1985; Pineau des For\^{e}ts, Flower,
\& Dalgarno 1988; Flower \& Pineau des For\^{e}ts 1994; Flower et 
al.\ 1996; Bergin, Neufeld, \& Melnick 1998).  When the simulations are
used to model the rich array of spectral observations now available for
molecular outflows, a combination of C- and J shocks
is broadly satisfactory, but an acceptable explanation for
important aspects of the observations remains elusive (e.g.,
Smith, Eisl\"{o}ffel, \& Davis 1998; Rosenthal, Bertoldi, \& Drapatz 
2000). Consequently, the theory of multifluid shocks remains of
considerable interest.

In this paper we reexamine a fundamental prediction of the theory,
namely, the ``critical speed,'' \vcrit, above which C type solutions
fail to exist. According to most prior work, the transition from C- to 
J-type solutions occurs when the neutral flow becomes too hot to remain
subsonic. For shocks in dense molecular gas, the transition
is caused either by the onset of \Htwo\ dissociation (which
decreases radiative cooling of the neutral gas) or by
self-ionization (which increases heating by ion-neutral scattering).
The precise value of \vcrit\ is sensitive to the rates
of various atomic and molecular processes.
DRD83 estimated that $49 \ge \vcrit \ge 40$\,\kms\ 
for preshock densities in the range
$10^4 \le \nH \le 10^6$\,\cmMMM.
Analogous calculations by Smith (1994) give similar
values, though recent simulations by Le~Bourlot et al.\ (2002)
predict somewhat larger critical speeds (e.g., 
$\vcrit \approx 70$\,\kms\ for $\nH=10^4$\,\cmMMM).
In any case, shocks in dense molecular
gas would be C~type for speeds $\gtrsim$40--50\,\kms\ if
the critical speed was determined solely by the temperature
of the neutrals.

However \vcrit\ is also constrained by the ``signal speed'' for 
compressive waves in the charged fluid, since shocks faster than the 
signal speed must necessarily be J type. In a cold, weakly-ionized 
molecular gas, $\vsig \simeq B/\sqrt{4\pi\rhoc}$, where $B$ is the 
magnetic field strength and \rhoc\ is the mass density of charged 
particles. If \rhoc\ is taken to be the density of atomic and molecular
ions, then $\vsig=\viA$, where $\viA\simgt 1000$\,\kms\ is the ion
\Alf\ speed. In this view the constraint imposed by the signal speed 
would be of no practical significance. However dust grains are also
(mostly) charged  (e.g., see Fig. 1 of Paper I) and, to the extent that
the grains are ``loaded'' onto the field lines, their contribution to
\rhoc\ also should be included (McKee, Chernoff, \& Hollenbach 1984).
For example, if \rhoc\ is taken to be the {\it total}\/
dust density, then $\vsig=\vgms$, where $\vgms\approx 5$\,\kms\
(see eq. [\ref{vgmsexpresseq}]) is the grain magnetosound speed.
A critical speed this small would have profound implications
for models of shocked molecular outflows. Flower \& Pineau des 
For\^{e}ts (2003) have computed the abundance of charged grains in
clouds including polycyclic aromatic hydrocarbons (PAHs) and 
cosmic-ray-induced UV photons. They suggest that only a small fraction
of the total grain mass in a plasma containing PAHs is charged and
coupled to the field, generally yielding magnetosound speeds
$\sim 50~\kms$. However, in the environments we wish to study, namely,
dense, dark clouds and cores, it is likely that most PAHs have condensed
or frozen onto the icy mantles of larger grains (Ehrenfreund \& Charnley
2000; Bernstein et al.  2003; Gudipati \& Allamondola 2003; Kr\"{u}gel
2003), and therefore should not affect the charge state of grains to 
such a large degree.

The critical speed $\vcrit$ would equal the grain magnetosound speed
$\vgms$ only if all the grains are well coupled to the magnetic field
lines (Chernoff 1987). However, the grains may be only weakly coupled
(e.g., Pilipp et al. 1990; Draine \& McKee 1993) and grain charge 
fluctuations limit the effective mass of the loaded grains to some 
fraction of the total dust mass (Ciolek \& Mouschovias 1993). 
Consequently, detailed numerical simulations are generally required to 
determine \vcrit. Although dust has been included in many detailed 
studies of shocks, most have neglected the inertia of the grains (i.e.,
the momentum equation for the dust has been omitted, on the assumption 
that the dust motion is force free). Because propagating waves in the 
grain fluid are suppressed when the grain inertia is neglected, studies
of this type cannot be used to examine the connection between grain 
loading and \vcrit. 

In this paper we present numerical calculations which
include the grain inertia, and extend Chernoff's work by
also including partial grain-field coupling and
grain charge fluctuations.
Our plan is as follows:
In \S~2 we summarize our assumptions and the free parameters
in our models.
The signal speed in a dusty plasma is discussed
in \S~3, where we examine the effects of dust on
the phase velocities and damping rates of compressive waves.
In \S~4 we present numerical solutions for steady and
time-dependent shock waves.
Our results are summarized in \S~5.

\vspace{-4ex}
\section{Modeling Assumptions and Parameters}
We simulate time-dependent, MHD shock
waves propagating in weakly-ionized plasmas using a finite difference
code which is described in detail elsewhere (Ciolek \& Roberge 2001,
henceforth Paper I).
We consider plane-parallel shocks
propagating along the $\pm z$ direction of a Cartesian
frame ($x$, $y$, $z$).
Fluid variables are assumed to depend only on $z$
and time $t$.
We restrict our discussion to ``perpendicular shocks'' with
a magnetic field of the form $\Bvec = B(z,t) \xhat$.
Fluid motions in the $x$-direction
(i.e., along the magnetic field)
are uncoupled from those in the $y$ and
$z$-directions and are ignored. Self gravity of the gas is also
neglected.

Our code models the plasma as a system of up to nine
interacting fluids: neutral particles
(denoted by subscript n), electrons (e), ions (i), large dust grains
(\gminus, \gzero, and \gplus) and very small grains or PAHs
(\sgminus, \sgzero, \sgplus) with charges $-1$, $0$, and $+1$
times the proton charge, $e$. The governing equations presented in 
Paper~I are solved here for the special case of vanishing PAH or
very small grain abundances since, as noted in the preceding section,
they are likely to be depleted onto large grains in dense cores.

Our code has the capacity to include radiative cooling of the neutral
gas by \Htwo, CO, and H$_2$O molecules, as well as cooling by 
collision-induced dissociation of \Htwo. For the calculations reported
in this paper, however, we ``turned off'' the neutral cooling. The
assumption of adiabatic flow was adopted for two reasons: First, the
structure of a {\it steady}\/ adiabatic shock with constant fractional
ionization can be calculated exactly (Chernoff 1987; Smith \& Brand
1990c), and this provides a valuable benchmark on our numerical results.
Second, Chernoff (1987) showed that shocks in a weakly-ionized
plasma undergo a C- to J-type transition at $\vcrit = 2.76\vnA$, where
$\vnA = B/\sqrt{4\pi\rhon}$ is the neutral \Alf\ speed.
This conclusion holds if (i) the charged particles are well coupled
to the magnetic field and (ii) the shock is subAlfv\'{e}nic in the charged
fluid.
Since these criteria are always satisfied by the ions,
the occurrence of a C-J transition at any other critical
speed, if found, can be attributed unambiguously to the effects of dust.
With neutral cooling turned off, only the abundances of the major
species H, \Htwo, and He (on which the ``source terms'' for
energy and momentum transfer between the charged and neutral
fluids depend; see Paper~I) are required.
We omit atomic hydrogen since the shocks considered here are
too slow to be dissociative and fix the other abundances
at $n\left(\Htwo\right)=0.5\nH$ and
$n\left({\rm He}\right)=0.1\nH$.

The ion fluid is composed of generic atomic (subscript \aplus)
and molecular (\mplus) ions with identical masses,
$\maplus=\mmplus=25$\,amu.
The ion abundances are described by a rudimentary chemical network
which allows for changes in the fractional ionization caused by
dissociative recombination, etc (see Paper~I).
In this paper we ``turned off'' the chemistry so that steady
solutions could be calculated exactly.
We neglected the inertia of the ions and electrons because they
attain force-free motion on length scales that are many orders
of magnitude smaller than the flow scale, and are therefore impractical
to track with our finite difference scheme.
Thermal pressure forces on the ions and electrons are
negligible compared to electromagnetic forces and gas drag, and so
were ignored.
Because the heat capacities of the ion and electron fluids are
small, their temperatures were calculated by requiring these fluids
to be in thermal balance (see Chernoff 1987).

The large grains (henceforth simply ``grains'') were assumed to be
spherical with identical radii, \ag, and to be composed of material
with density 3\,\gcmMMM.
Realistic simulations with a spectrum of grain sizes will
be presented elsewhere.
Our purpose here is to elucidate the essential physics.

The free parameters in our calculations are listed in Table~1.
Parameters that describe the undisturbed preshock plasma
have realistic values for a typical cloud core, except that
the cosmic ray ionization rate has been artificially adjusted
so that all calculations have the same fractional ionization.
We considered two simple dust models, in which the dust
is either weakly coupled (the WC model) or strongly coupled
(the SC model) to the magnetic field.
Both models have a realistic  dust-to-gas mass ratio,
but the grain radii were chosen artificially to give
large and small values for the grain Hall parameter,
\be
\label{defHallgeq}
\Hallg = \Omegag \tdraggn~,
\ee
where $\Omegag$ is the charged-grain gyrofrequency and $\tdraggn$ is the
grain-neutral collision time. We give values of \Hallg\ and other 
pertinent quantities in Table~2.

\section{Compressive Disturbances in a Weakly Ionized, Dusty Plasma}
\vspace{-1ex}
\subsection{Wave Modes}
A weakly-ionized, dusty plasma supports a variety of waves with phase
velocities and damping times that can be obtained by solving the 
appropriate dispersion relations. Pilipp et al.\ (1987, hereafter 
PHHM87) solved the dispersion relation for \Alf\ (i.e., noncompressive)
waves including the effects of partial field coupling and charge 
fluctuations. PHHM87 found that, when the grains are well coupled to the
field lines, their added inertia reduces the phase velocity relative
to a dust-free plasma that is otherwise identical. Ciolek \& Mouschovias
(1989, hereafter CM89) carried out the analogous calculations for waves
propagating at an arbitrary direction relative to the field.
Here we consider the special case of CM89's analysis for the
magnetosonic (i.e., compressive) modes relevant to the formation
of magnetic precursors. Since some of the physics is both interesting 
and not published elsewhere, we describe the results in some detail.

We obtained the dispersion relation by linearizing the equations
of motion about a uniform, static unperturbed state
and searching for monochromatic plane-wave solutions of the form
\be
q_{1}(z,t) = \tilde{q}_1(k)
   \,\exp\left[\imath\left(kz - \omega t\right)\right],
\ee
for the first-order perturbation $q_1$ in variable $q$.
The perturbations were assumed to have the geometry appropriate
for magnetosonic modes.
We linearized the
equations of motion from Paper~I with several modifications:
We omitted the equations of motion for very small grains
on the assumption they are depleted (see \S2).
We omitted the continuity equations for the ions and grains
because, for the very low fractional ionizations of interest here,
their density perturbations have virtually no affect on the dynamics.
We included the momentum equation for the ions (the ion inertia
is neglected in Paper~I and in our numerical code)
since otherwise waves in the ion fluid would be suppressed.
We assumed for simplicity that the perturbations are adiabatic, and
therefore omitted the energy equation for the neutral fluid.
Finally, in writing the magnetic induction equation we assumed
that magnetic flux is frozen into the electron fluid.
This is more general than the corresponding assumption in Paper~I
(flux freezing in the ion fluid) though equivalent in the limit
of vanishing small grain abundance.
Our results for the WC and SC models are shown in Figures~1 and 2,
respectively, where $\vphi\equiv \real[\omega]/k$
is the phase velocity and $\tdamp \equiv -1/\imag[\omega]$
is the damping time. Several wave modes are apparent.

\subsubsection{Ion Magnetosound Waves and Quasiparticle Oscillations}
At very short wavelengths only the ions and electrons
are coupled to the magnetic field.
In the ion magnetosound mode (labeled ``ims'' in Fig.~1-2),
the ions, electrons and magnetic field
oscillate against a background of fixed neutrals and grains.
The phase velocity is the ion magnetosound speed,
\be
\vims \simeq \viA = {B \over \sqrt{4 \pi \rho_{\rm i}} },
\label{vimsdef}
\ee
where $\vims\simeq \viA$ because
we have neglected thermal pressure of the ions.
For the core conditions adopted here, $\vims \sim 1000$\,\kms\ (Table~3).

Ion magnetosound waves do not exist for wavelengths larger
than some value \lambdaims.
In the WC model, \lambdaims\ is the
familiar cutoff imposed by ion-neutral friction,
\be
\label{lambdaimseqa}
\lambdaims = 4 \pi \vims \tdragin
~~~~~~~~~\mbox{(WC model)}
\label{msimaxlameq}
\ee
(e.g., Kulsrud \& Pearce 1969),where \tdragin\ is the ion-neutral
drag time.\footnote{In general, \tdragij\ denotes the drag time for particles
of type i to be slowed by elastic scattering with particles of type j,
and \tdragitot\ denotes the drag time for particles of type i to
be slowed by scattering with all other particle types.}
In the SC model, the cutoff for ion magnetosound waves is
the {\em electrostatic wavelength},
\be
\lambdaims =
\lambdael \simeq \frac{2 \pi \vims}{\Omegaqp}
~~~~~~~~~\mbox{(SC model)}.
\label{electrolambda}
\ee
For $\lambda\ge\lambdael$ the ion magnetosound mode transforms into
a nonpropagating\footnote{Note that $d(\log \vphi)/d (\log \lambda) = 1$
on the branch labeled ``qp''.}
``quasiparticle'' mode (labeled ``qp'' in
Fig.~1--2).\footnote{The quasiparticle mode also appears in the WC
model (Fig.~1). However, quasiparticle effects are insignificant in the
WC model: the abundance of charged grains, and hence \Qeff,
are $\approx 60$ times smaller than in the SC model
(see Table~2). As a result, \Omegaqp\ is so small that the
quasiparticle oscillations are evanescent.}
In the presence of charged grains, the ions and electrons
collectively form a single ``quasiparticle fluid'' of
electron-shielded ions with effective charge
\be
\Qeff= \left( {\nni - \nne \over \nne} \right) e
     \simeq \left({\ngm \over \nne}\right) e.
\ee
The ions and electrons gyrate about the field lines
at a constant frequency
\be
\label{OmegaQdefeq}
\Omegaqp = \left(\frac{\nni - \nne}{\nne}\right) \Omegi
= \left(\frac{\ngm - \ngp}{\nne}\right) \Omegi
\simeq \left(\frac{\ngm}{\nne}\right) \Omegi,
\ee
where $\Omegi = eB/\mi c$ is the ion gyrofrequency.
For detailed discussions on the physics of quasiparticle oscillations,
see Ciolek \& Mouschovias (1993, 1994, 1995).

In both models, propagating ion magnetosound waves are damped
by ion-neutral drag on a time scale
\be
\taudims = 2 \tdragin,
~~~~~~~~~\mbox{($\lambda<\lambdaims$)},
\label{taudimsdef}
\ee
where the factor of 2 occurs because
of equipartition between the kinetic and magnetic field energies.
For $\lambda > \lambdaims$ ion-magnetosound waves are evanescent:
the ions and magnetic field diffuse through the rest of the plasma
with damping on the diffusion time scale,
\be
\label{iondifftime}
\taudims = {\lambda^2 \over \pi \vims \lambdaims}
~~~~~~~~~\mbox{($\lambda>\lambdaims$)}.
\ee

\subsubsection{Grain Magnetosound Waves}
At somewhat larger wavelengths, grains participate in
the wave motions provided they are coupled to the magnetic field.
Thus, grain magnetosound waves (labeled ``gms'') appear in the SC model
but not in the WC model. In the former, $\gminus$ and $\gplus$ grains 
are directly attached to the field lines and the neutral grains are 
effectively attached by charge fluctuations.\footnote{Attachment of the
neutral grains occurs if the time for a neutral grain to capture an
electron or ion is much less than the $\gzero$-neutral drag time,
\tdraggzn\ (PHHM87; CM89; Ciolek \& Mouschovias 1993). This inequality
is satisfied by the SC and ``intermediately-coupled" IC (see \S~4.2.1)
models.}
Accordingly, ions, electrons and dust all contribute
to the inertia of the wave and
the phase velocity is
the grain magnetosound speed,
\be
\label{vgmsdefeq}
\vgms =
{
B
\over
\sqrt{4\pi \left(\rho_{\rm i}+\rho_{\rm e} + \rho_{\rm g}\right)}
},
\label{vgmsdef}
\ee
where \rhog\ is the total density of charged and uncharged dust.
Because the dust density is much larger than the ion density,
the grain magnetosound speed is much smaller than the ion
magnetosound speed ($\vgms \ll \vims$, see Table 3).

Grain magnetosound waves are allowed for wavelengths larger than
\begin{equation}
\label{mingraincutoffeq}
\lambdamingms \simeq
 \frac{\pi \vgmms}{\Omegaqp \tdragin \cdot \Omegag},
\end{equation}
where \vgmms\ ($\simeq \vgms$, since
$n_{{\rm g^{-}}} \gg n_{{\rm g^{0}}},~n_{{\rm g^{+}}}$
--- see Table 2) is the magnetosound speed in the fluid
of {\em charged}\/ grains
and $\Omega_{\rm g}$ is the grain gyrofrequency.
The minimum wavelength $\lambdamingms$ signifies
some interesting and unique plasma physics: propagation of magnetosound
waves in the grain fluids is aided by an electrostatic attraction
between the positively-charged quasiparticles and the negatively charged
grains (CM89). For $\lambda \ge \lambdamingms$, the electrostatic
attraction between the quasiparticles and charged grains allows the
charged grains to be ``dragged along'' with the magnetically attached 
quasiparticles and the field (see Ciolek \& Mouschovias 1993, \S~3.1.2).
The maximum wavelength for grain magnetosound waves is determined
by gas-grain friction.  The cutoff occurs at
\be
\label{msgmaxlameq}
\lambdamaxgms =
{
4 \pi \vgms \tdraggn
\over
1 + \left[
\left(\rhoi/\rhog\right)
\left(\tdraggn/\tdragin\right)
\right]
}
\ee
(CM89).
For $\lambda > \lambdamaxgms$, the waves are evanescent and
the combined plasma of ions, electrons and dust diffuses relative to
neutrals.

At small wavelengths (i.e., $\lambda \sim \lambdamingms$),
grain magnetosound waves are damped primarily by the drift of
quasiparticles and magnetic field relative to the charged grains.
The damping time is
\be
\label{quasidampeq}
\taudgms \simeq  \frac{1}{2 \pi}~\frac{\lambda^{2}}{\lambdamingms~\vgmms}
~~~~~~~~~~~~ \mbox{if}\ \lambda \sim \lambdamingms
\ee
and
this ``damping branch'' therefore has a slope
$d \log \tdamp / d \log \lambda = 2$ in Fig.~$2b$.
At longer wavelengths the diffusion of quasiparticles with respect
to grains is significantly lessened and the waves are damped by the
frictional drag on the grains and ions exerted by the
neutrals (which are essentially motionless
at these frequencies).
In this regime the damping timescale becomes
independent of $\lambda$ and the damping curve in Fig.~$2b$
flattens.
The damping rate is the sum of the damping rates
for ion-neutral and grain-neutral friction, so that
\be
\label{grwavedampeq}
\taudgms \simeq 2 \left\{
\left[\frac{\rhoi/\rhog}{\tdragin}\right]
+ \left[\frac{1}{\tdraggn}\right]\right\}^{-1}
~~~~~~~~~~~~ \mbox{if}\ \lambdamingms \ll \lambda \le \lambdamaxgms,
\ee
where
\be
\label{totgrdampeq}
\frac{1}{\tdraggn}
\equiv
\sum_{\delta}
\frac{\rho_{\rm g^{\delta}}}
{\rhog~\tau_{\rm drag}^{\rm g^{\delta}, n}}.
\ee
and the factor of 2 again appears because of
equipartition between the kinetic and magnetic field energies in the wave.
For $\lambda > \lambdamaxgms$ the (evanescent) waves are
damped on the diffusion time, with
\be
\label{taudgmsev}
\taudgms \simeq
{ \lambda^{2} \over \vgms~\lambdamaxgms}
~~~~~~ \mbox{if}\ \lambda > \lambdamaxgms.
\ee

\subsubsection{Neutral Sound and Magnetosound Waves}

Oscillations of the neutral fluid occur for all wavelengths.
At small wavelengths, adiabatic neutral sound waves (labeled ``ns'')
occur wherein the neutrals oscillate in a sea of fixed ions and grains.
The phase velocity is the adiabatic sound speed,
\be
\vnas = \gamma^{1/2}\Ciso \simeq \left(
{5 \kB T_{\rm n} \over 3m_{\rm n}}
\right)^{1/2},
\ee
where $\gamma$ is the adiabatic index of the gas
(Table~3).
Sound waves in the neutral fluid are damped by collisions
between neutrals and the other fluids, so that
\be
\label{sounddampeq}
\taudns = 2
\left[
\sum_{\alpha}
\left(
\frac{1}{
\tau^{\rm d}_{{\rm n,\alpha}}
}
\right)
\right]^{-1}.
\label{totdragntimeeq}
\ee

At longer wavelengths the ions and grains
excite neutral magnetosound waves (labeled ``nms'') consisting of
longitudinal compressions of the bulk fluid
(i.e., the neutrals plus the electron-ion-grain plasma)
and magnetic field.
The phase velocity is the neutral magnetosound speed,
\be
\vnms = \left( C_{\rm ad}^2 + V_{\rm nA}^2\right)^{1/2},
\ee
(Table~3).
Since the neutral fluid density far exceeds the combined densities of
all other fluids, the inertial response of the wave is
almost entirely due to the neutrals.
Thermal-pressure and magnetic forces --- the latter
affecting
the neutral fluid indirectly through collisions between neutrals
and charged particles --- provide the restoring force required for
wave propagation.

Propagation of neutral magnetosound waves
becomes possible for wavelengths greater than
the {\em minimum neutral magnetosound wavelength},
\be
\lambdaminnms = \pi \left(\frac{V_{\rm nA}^2}{\vnms}\right) \tdragntot
\label{nmscutoffeq}
\ee
(CM89).
Note that for $\vnA \gg \vnas$,
$(v_{\rm nA}^{2}/\vnms) \rightarrow \vnA$.
The minimum neutral magnetosound wavelength is the perpendicular
($\kvec \perp \Bvec$) analog to the lower cutoff wavelength for neutral
\Alf\ waves ($\kvec \parallel \Bvec$) in ion-neutral (Kulsrud \& Pearce
1969) and dusty (PHHM87)  plasmas. The damping of neutral magnetosound
waves is due to ambipolar diffusion of the plasma and magnetic field 
with respect to the neutrals, with 
\be
\taudnms = {\lambda^{2} \over 2 \pi^{2} V_{\rm nA}^2 \tdragntot}
\label{taudnms}
\ee
(see Fig.~$1b$,$2b$). For $\lambda > \lambdaminnms$ there also exists a
non-propagating ($|\vphi| = 0$) thermal-pressure-driven diffusion mode,
in which the neutrals diffuse relative to a fixed background of plasma
and magnetic field, which has
$\tdamp = \lambda^{2}/4 \pi^{2}\Ciso^2 \tdragntot$.

\subsection{Propagation of Small-Amplitude Disturbances}

The preceding section shows that disturbances in
a dusty plasma can excite a variety of waves, with phase
velocities and damping rates that span several orders of magnitude.
We are ultimately interested in how compressive  waves in the charged
fluids affect the formation of C shocks, and we do this in \S4 by
simulating large-amplitude disturbances (i.e., shocks).
However, it is instructive to consider first a simpler problem,
the propagation of small-amplitude, Gaussian wave packets,
where interpretation of the dynamics in terms of different
modes is much simpler:
the narrow range of length scales in such a packet
limits the number of modes that are excited, and the linearity
of the disturbance allows us to interpret the results using \S3.1.

At time $t=0$ we superimposed a perturbation
\begin{equation}
\label{Gaussformeq}
B(z) = \delta B \exp\left[-\frac{(z - Z/2)^{2}}{\ell_B^2}\right]
\end{equation}
on a static, uniform initial state corresponding to either the
WC or SC model.
We centered the pulse initially on a computational domain
$\left[0,Z\right]$ and followed its evolution with our numerical
code.
The code uses transmissive boundary conditions, 
which allow the pulse to exit the domain freely, i.e., without
generating reflected pulses at the boundaries.
Thus, the pulse should evolve exactly like an identical pulse
in an infinite computational domain.
In all cases we set $\delta B/B_0=0.1$ but chose a few
different values of the packet width, \ellb.

Simulations of a wave packet with $\ellb=4\times 10^{14}$~\cm\ are
described in Figure~3 for the WC model, where
the left-hand panel shows fluid velocities and the
right-hand panel shows the density and magnetic field.
The upper and lower boxes show variables at times
$t_1=8.5$~yr and $t_2=17$~yr, respectively.
These times are so short ($t_1,t_2\ll \tdragni$, see Table~2)
that the neutrals are virtually stationary.
The grains, which receive their ``marching orders'' in the WC model
mainly by colliding with the neutrals, are also motionless.
The ions are not stationary, but the initial disturbance does
not propagate: the ``wave'' packet simply sits at the center of the
grid and decays away. The ions in Fig.~3 are diffusing through the 
grains and neutrals at the terminal drift speed determined by the
balance between ion-neutral drag and the driving magnetic pressure 
gradient. The absence of wavelike (i.e., oscillatory) motions
may seem surprising, since propagating ion magnetosound waves
are permitted by the dispersion relation on the length scales
$\sim \ellb$ in the initial pulse (Fig.~$1a$).
However the damping time for propagating waves in the ions is so
short ($\sim 0.01$~yr, see Fig.~$1b$) that such waves have virtually
no effect on the dynamics.\footnote{Of course, ion magnetosound
waves are suppressed by our numerical code, where the ion inertia
is neglected. The smallness of the damping time compared to other
dynamical time scales of interest justifies this omission.}
We conclude that, on time scales of practical interest,
ion motions in a compressive disturbance are the diffusive motions
associated with evanescent waves.

Figure~4 describes the same initial disturbance as Figure~3 
but for the SC model where grain magnetosound waves are
allowed for wavelengths $\sim\ellb$ (Fig.~$2a$).
In contrast to the preceding example, the initial disturbance
gives rise to two symmetric wave packets propagating in
the $\pm z$ directions.
That the pulse in Fig.~4 is indeed a  packet of grain magnetosound waves
can be confirmed by computing the propagation speed:
the centers of the forward- and backward-propagating
pulses travel a distance $\simeq 1.2 \times 10^{15}~\cm$
between $t_1=8.5$~yr (Fig.~$4b$) and $t_2=77$~\yr\ (Fig.~$4d$).
The propagation speed is therefore $5.0~\kms$, in excellent agreement
with \vgms\ (Table~3).
The packet is also damped at the expected rate.
Taking the width of the pulses ($\simeq 2.6 \times 10^{15}~\cm$)
at time $t_2$ as a characteristic
wavelength, we find from Fig.~$2b$ that the 
damping time should be $\simeq 92~\yr$.
Since the initial amplitude of each counterpropagating pulse is $ 0.05 B_0$,
the value of $B/B_0$ at the peak of the pulse should therefore
be $\simeq 1 + 0.05\exp(-77/90) = 1.022$ at $t_{2}$.
This is very close to the numerical value, 1.024, for the data in Figure $4d$.
Finally, the linear analysis for grain magnetosound waves predicts that
the magnitude of
the grain velocity\footnote{The neutral and charged grains behave
as a single fluid in our simulations of wave packets
because the charge fluctuation time
is small compared to the flow time scale ($\sim \ellb/v$).}
should obey the relation
\begin{equation}
\label{linveloceq}
\frac{|v_{{\rm g}}|}{\vnis} \simeq
\left(\frac{\vgms}{\vnis}\right)\left[\left(\frac{|B|}{B_0}\right) - 1\right]
\end{equation}
in the limit of weak damping,
where \vnis\ is the isothermal sound speed (Table~3).
At points near the peaks of the pulses, the numerical
data for Fig.~$4c$ and $4d$ satisfy this relation to better than
one percent.

Figure~4 also illustrates the different responses of the various
fluids to the compressive disturbance.
For example, the neutrals are unable to respond at
the  frequencies corresponding to length scales $\sim\ellb$
and act simply as a stationary background.
The grain velocity lags somewhat behind the
ion velocity (Fig.~$4a$,$c$) because the time for the
charged grains to respond to the magnetic field variations
($\Omega_{\rm g}^{-1}$=63~\yr, see Table~2) is not much smaller
than the flow time scale ($\simlt 10^{15}\cm/5~\kms = 65~\yr$).
In fact the grains repond somewhat faster than this
because of their attraction to the electron-shielded ion quasiparticles,
which are attached to the magnetic field.
Thus, the ion and grain motions are becoming synchronous by
time $t_{2}=72~\yr$ (Fig.~$4c$).

Figure~5 describes another Gaussian packet in the SC model, but
with a larger width, $\ellb=1.1\times 10^{16}$, that
is close to the maximum wavelength for propagating grain
magnetosound waves (Fig.~$1a$).
As expected, the motion is largely diffusion of the ion-grain plasma
with respect to the neutrals, with some additional effects caused
by the presence of Fourier components at wavelengths short enough
for propagating grain magnetosound waves. Thus, the magnetic field pulse
is more smeared out than the analogous pulse in Fig.~3, and has
``rounded shoulders" (compare Figs.~$3d$ and $5d$). The damping time
scale ($\sim 330~\yr$, see Fig.~$1b$) for the wave modes is much
less than the decay time ($\simgt 5 \times 10^{3}~\yr$) for the 
longer-wavelength diffusion modes. Grain magnetosound waves are 
therefore overwhelmingly damped before wave packets can ``break out" of 
the initial Gaussian pulse. After times comparable to the response time
of the neutrals ($\sim\tdragng$, see Table~2) the neutrals acquire small
nonzero velocities (Fig.~$5c$) away from the center of the initial 
pulse, where a density depression appears (Fig.~$5d$).

Figure~6 describes an even wider packet
with $\ellb = 2 \times 10^{17}~\cm$ in the SC model.
The width now exceeds the minimum wavelength for
(propagating) {\em neutral}\/
magnetosonic waves (Fig.~$1a$).
At times shorter than $\sim\tdragng$ (Table 3), the velocity of the 
neutrals lags that of the ion-grain fluid (Fig.~$6a$,$b$).
At much later times (Fig.~$6c$,$d$), the ions, grains, and neutrals
behave as a single fluid and packets of neutral magnetosound waves start
to propagate in the $\pm z$ directions. Because the neutrals are 
effectively loaded onto the magnetic field lines on these scales, the
signal speed is greatly reduced. The data for Fig.~6 imply a propagation
speed of about $0.57$~\kms, close to the neutral magnetosound speed 
(Table~3).


\section{Simulations of Multifluid Shock Waves with Dust}

We now consider shock waves, which were simulated as follows.
At an initial time, $t=0$, the flow was assumed to be a J shock
propagating in the $+z$ direction, with the jump front located
at the left-hand boundary of the domain, $z=0$.
The preshock ($z>0$) region was assumed to be uniform and static initially
with values of the density, magnetic field strength and other
variables taken from either the WC or SC model.
Initial values of the post-shock
($z<0$) variables were calculated by choosing some initial
shock speed, \vsi, and applying the adiabatic jump conditions (assuming
a frozen-in magnetic field) at $z=0$.
The flow at later times was determined using our numerical code.
Transmissive boundary conditions were applied at both
boundaries to simulate a domain extending to $z=\pm\infty$.
Adiabatic flow was assumed and the chemistry was turned off;
the effects of cooling and chemistry will be
discussed in later papers.

\subsection{Dust and the Structure of Steady Shocks}

We computed a sequence of steady solutions for increasing shock
speed, \vs, by integrating the equations of motion to a steady state.
For each speed we computed solutions for both the WC
and SC models in order to compare effects of dust in the two cases.
Instead of the shock speed, it is useful to characterise the steady
solutions by the neutral \Alf\ Mach number,
\be
\Mna \equiv {\vs \over \vnA},
\ee
or the grain \Alf\ Mach number,
\be
\Mga \equiv {\vs \over \vgA},
\ee
where $\vgA=\vgms$ is the grain \Alf\ speed. Recall that for a 
two-fluid (ion+neutral) plasma adiabatic shocks have $\vcrit=2.76\vnA$
(Chernoff 1987) or $\vcrit=1.38$\,\kms\ for the \Alf\ speed adopted here
(Table~3). To the extent that our solutions resemble the two-fluid case,
we expect C~type solutions only for shock speeds much smaller than the 
speeds ($\simgt 10$\,\kms) in shocked molecular outflows. Thus, our 
numerical results bear little resemblance to real shocks: dust physics 
is the issue of interest here.

In fact we expect the WC solutions to be virtually identical to
two-fluid shocks with no dust: except for regions very close to a jump 
front, the grains should move with the neutrals, and therefore have no 
role in momentum or energy transfer between the charged and neutral
fluids. This is verified by Figure~7. The symbols show velocities 
computed numerically for a shock with $\Mna=2.1$ using preshock 
conditions from the WC model. The smooth curves are the exact solution 
for a steady, two-fluid shock of the same speed (Chernoff 1987; Smith \&
Brand 1990c).

Figures 8--11 describe a sequence of steady shocks with
\Alf\ Mach numbers increasing from $\Mna=2.1$ to $\Mna=5.0$.
It is difficult to pinpoint the C-J transition in these
solutions because J-shocks are weak near
the transition point (the compression ratio of the shock is
unity precisely at the C-J transition) and because artificial
viscosity smears out the J front.
However it is apparent from Figs.~8--11 that (i) the transition
is at or near $\Mna=2.76$ for the WC model; and
(ii) the transition occurs at or near the {\em same}\/
 \Mna\ value for the SC model. The first result is expected for reasons
noted above. The second result occurs because the critical speed,
$\vcrit\simeq 1.38$\,\kms, is smaller than the grain magnetosound speed,
$\vgms=5$\,\kms.  That is, the transition to J-type flow occurs because
the neutrals become too hot, not because the shock speed exceeds
any signal speed.

The principal difference between the WC and SC solutions in
Fig.~8--11 is that the magnetic precursor is thinner in the SC models.
Draine (1980) pointed out that the width of the precursor is
is inversely proportional to the rate per unit volume, $F$, at
which momentum is exchanged between the charged and neutral fluids.
In the WC models momentum transfer is due almost entirely to
ion-neutral scattering.
Dust contributes virtually nothing because the grains are
almost at rest in the neutral fluid.
However,
in the SC model grain-neutral scattering actually contributes
somewhat more to $F$ than ion-neutral scattering
(note that $\tdragng<\tdragni$ in
Table~2), and this makes the precursors thinner.

We conclude that the signal speed in the grain fluid has no
effect on the C-J transition for adiabatic shocks if
$\vgms>2.76\Mna$.
We say that C shocks are ``cooling limited'' in this regime.
To explore the regime $\vgms<2.76\Mna$, we constructed a ``heavily
loaded'' (HL) grain model identical to the SC model (i.e., the grains
are well coupled to the field) but with an extremely large dust-to-gas
mass ratio, $\rhog/\rhon=0.5$, chosen to make $\vgms=0.71$\,\kms.
Figure~12 shows a sequence of steady shocks with increasing
speeds computed for the HL model. The solutions are C type for $\Mga<1$
and J type for $\Mga>1$, as predicted by Chernoff (1985). We say that C
shocks are ``signal limited'' in this regime, since the C-J transition
occurs when \vs\ exceeds the signal speed (=\vgms) in the charged fluid.
Notice that the ions are forced to move with the grains in these models,
since both are coupled to the magnetic field. Consequently there is no
magnetic precursor in any fluid for shocks with $\Mga>1$.\footnote{The
slight offsets in Fig.~$12c$, $d$ between the velocities of charged and
neutral particles is a numerical artifact: they are only one grid zone
wide.}

\subsection{Dust and the Formation of Magnetic Precursors}

Our study of small-amplitude disturbances shows
that the response of charged particles to
a short-wavelength disturbance depends dramatically
on whether or not the dust is coupled to the magnetic field.
It is therefore of interest to study how a magnetic precursor
forms in these two different cases.
Figures~13 through 15 compare shocks in the WC and SC
models for a sequence of increasing times.
Except for the grain models, both calculations started with
identical initial conditions: J shocks with identical
compression ratios at $z=0$ and uniform, static plasma for $z>0$.
The {\em steady solutions}\/ for the SC (Fig.~$15b$) and WC (Fig.~$15c$)
models are similar, consistent with the fact that the steady state
has a shock speed $\vs<2.76\vnA$ (see \S4.1).
However, the intermediate solutions are dramatically different.

Our linear analysis (\S3.1) suggests that differences in the
flows might be caused by the different mechanisms for
transmitting compressive disturbances in the two cases.
Strictly speaking the disturbances in Fig.~13--15 are nonlinear;
however the linear analysis should give a reasonable description
of the charged fluids, since $\vi\ll\vims$ and $\vg\ll\vgms$.
We therefore adopt the view that the discontinuity imposed
by the initial J~shock ``excites''
modes with all possible wavelengths initially, and interpret
the subsequent evolution of the flow in terms of these different modes.

Consider first the WC solution. The earliest times shown (Fig.~$13a,c$
and Fig.~14a) are much smaller than the dynamical response times
of the dust ($\Omega_{\rm g}^{-1} \sim 4100$\,yr) and the neutrals
($\tdragntot\sim 10^4$\,yr) so only the modes corresponding to pure ion
motions are initially of interest. Since these times are also much
{\it longer}\/ than the damping time for propagating ion magnetosound
waves (which are suppressed by our code in any case), the rapid ion
motions in Fig.~$13a,c$ correspond to evanescent ion magnetosound waves.
To see that this interpretation is correct, recall (\S3.1) that
evanescent waves correspond to diffusive motions. A linear analysis of
the ion equations of motion (Appendix~A) shows that the diffusion 
coefficient for the ions is
\be
\label{difficoefeq}
{\cal D}_{\rm ion} = 4\pi^2\vims^2\tdragin.
\ee
After a time $t$ we therefore expect the ``front'' separating
accelerated from stationary ions to have propagated a distance
\be
\label{zfronteq}
z_{\rm front}(t) = \left({\cal D}_{\rm ion} t\right)^{1/2} 
\ee
upstream from the initial J shock. In fact, the predictions of 
eq.~(\ref{zfronteq}) are in remarkably good agreement with the numerical
solution for the WC model (Fig.~16). We conclude that a magnetic 
precursor forms via ion diffusion when dust is dynamically unimportant.

For early times (Fig.~$13a,c$), the ions simply diffuse through an 
almost-stationary background of neutrals and dust. After a few hundred
years, these ion motions have compressed the magnetic field sufficiently
to make the dynamical response time of the grains 
($\sim \Omega_{\rm g}^{-1}\propto B^{-1}$)
less than the flow time. At this point (Fig.~$14a$) betatron 
acceleration (Spitzer 1976) starts to drive the grain velocity toward 
the ion velocity. After another few thousand years the neutrals have 
responded significantly (Fig.~$14c$). Thereafter (Fig.~$15a,c$) the 
evolution slows as the charged and neutral fluids approach mechanical 
equilibrium, with magnetic pressure balanced by neutral drag in the 
charged fluid and ion-neutral drag balanced by thermal pressure in the
neutral fluid.

Our conclusion --- that a magnetic precursor forms via ion diffusion 
rather than wave propagation when the dust is dynamically unimportant---
has important implications. Evidently the conventional wisdom, that the
ion magnetosound speed sets an upper limit on \vcrit\ when neutral 
cooling is strong, is incorrect. The upper limit in this situation must
occur when the diffusion time scale becomes smaller than the time for
ion-neutral scattering to accelerate the neutral gas. An important 
objective of our future work is to determine whether this upper limit 
constrains real shocks with a spectrum of dust sizes and molecular 
cooling.

Now consider the SC solution. Comparing $z_{\rm front}(t)$  for the SC 
and WC models (Fig.~16), we see that the dynamics of the magnetic 
precursor are {\em qualitatively different} in the two cases. At very 
early times there is evidence for ``$t^{1/2}$ behavior'' in the SC model
(e.g., the curve in Fig.~16 is concave down for small times and the 
slope is very large near $t=0$). We attribute this early time behavior 
to diffusive motions in the ion fluid, which is not strongly coupled to 
the grains for $t \ll \Omega_{\rm g}^{-1}$. At later times,
$z_{\rm front} \approx \vgms t$ is a good approximation for the SC
model. Clearly, the magnetic precursor forms in the SC model via the
propagation of grain magnetosound waves. Now there is a well-defined 
signal speed, \vgms, and it is correct to say that $\vcrit=\vgms$.
Of course, this is precisely what we found in Fig.~12.

It is of interest to note that a steady C shock forms at an
earlier time in the SC model ($\approx 5 \times 10^4$ yr, Fig.
$15a$) than in the WC model ($\approx 2.6 \times 10^5$ yr, Fig.
$15c$). This can be understood for the following reasons: First, in the
SC model, ions and grains move together and collide in parallel (i.e.,
simultaneously) with the neutrals. This gives a net collision or 
momentum-exchange time 
$\tau^{\rm (n,tot)}_{\rm drag,SC} = \tdragng (1 + \tdragni/\tdragng)^{-1}$
(e.g., see eq. [\ref{sounddampeq}]). For the WC model, only 
collisional drag between ions and neutrals is important, and 
$\tau^{\rm (n,tot)}_{\rm drag,WC} = \tdragni$.
Inserting the values listed in Table 2, we have
$\tau^{\rm (n,tot)}_{\rm drag,SC}/\tau^{\rm (n,tot)}_{\rm drag,WC} = 0.4$.
Hence, the collisional effect of the ions and grains acting in concert
results in momentum transfer that more rapidly ``sweeps up" the
pre-shocked neutrals in the SC model compared to that which occurs in
the WC model. Secondly, the driving mean magnetic field gradient in the 
precursor region in the SC model is greater than in the WC model. As
discussed above, this is due to the difference in the precursor 
formation mechanism between these models. Rapid ion diffusion disperses
the magnetic field ahead of the shock in the WC model over a much larger
region than in the SC model (Figs. $15a$, $c$).
Because the mean field gradient on the plasma is greater in 
the SC model than in the WC model, the neutrals thereby also experience
a greater accelerating frictional force due to the larger gradient in
plasma velocities occurring over a much shorter lengthscale. 
Thus, taken together, the combined effects of greater momentum transfer 
by collisions in a more narrowly confined magnetic precursor allows a
steady C shock to form at a significantly earlier time in the SC model
than in the WC model.
\vspace{-5ex}
\subsubsection{Discussion}
For the SC model, which has a population of single-size grains with 
radius $\ag = 0.05~\mu{\rm m}$ and Hall parameter $\Hallg = 4.2$, the
grain magnetosound speed (eq.  [\ref{vgmsdefeq}]) has the typical value
\be
\label{vgmsexpresseq}
\vgms \simeq 5 \left(B\over{50~\mu{\rm G}}\right)
\left(2 \times 10^{4}~\cc\over{\nn}\right)^{1/2}
\left[\frac{10^{-2}}{(\rhog/\rhon)}\right]^{1/2}~\kms .
\ee
However, the single-size grain model that we have assumed here is an 
idealization to the interstellar medium; in actuality, grains have a 
distribution in grain sizes, normally spanning more than several orders
of magnitude in radii. The very large grains in such a 
distribution will have $\Hallg < 1$ and will therefore not be coupled to
the magnetic field (as in the WC model), while the smaller
grains will instead have $\Hallg > 1$ and will be attached to the field
(as in the SC model). The relevant question is then, what is
the correct dust abundance that should be used to calculate the
numerical value of $\vgms$ when there is a distribution in grain sizes?

A likely answer to this question is that one should use the fraction of
the charged dust
\footnote{It is assumed here that a non-negligible fraction of the 
grains are charged (see Table 2); this would also encompass grain models
containing a population of very small grains for densities 
$\simlt 10^{7}~\cc$ (e.g., Nishi, Nakano, \& Umebayashi 1991; see,
also Fig. 1 of Paper I). As noted in \S~1, rapid charge fluctuation and 
momentum transport between neutral and charged grain fluids can 
effectively couple even the neutral grains to the magnetic field 
(PHHM87; CM89; Ciolek \& Mouschovias 1993). If, however, there is still
an extant, non-negligible population of PAHs at these densities, the 
overall fraction of charged grains may be greatly reduced, with a 
corresponding increase in $\vgms$ (Flower \& Pineau des For\^{e}ts
(2003).} 
in the distribution that is coupled to the magnetic
field. Neglecting grain-quasiparticle
interactions, this corresponds to grains with $\Hallg \simgt 1$.
That the criterion $\Hallg \simeq 1$ is the proper one for determining
whether charged grains are attached to the magnetic field can be
seen in Figures $17a$, $b$, which show C shock results for two
``intermediately coupled" models, IC-1 and IC-2, respectively.
The two models have the same physical parameters as in the WC and
SC models (including ion abundances), except that they contain
different sized grains: IC-1 has grains with $\ag = 0.125~\mu{\rm m}$
and $\Hallg = 1.1$, while IC-2 has grains with $\ag = 0.09~\mu{\rm m}$
and $\Hallg = 1.3$. The initial conditions for the shock flow are
the same as for the SC and WC models in Figs. 13 - 15. As in the
other models, a steady C shock eventually forms: in IC-1 this occurs at
$1.70 \times 10^{5}$ yr, while in  IC-2 it is at 
$1.25 \times 10^{5}$ yr. Comparison of these models with the WC and
SC models (Figs. $15a$, $c$) shows that, even though the coupling of
the grains to the magnetic field in the IC-1 and IC-2 models is not 
perfect, the shock structure has been significantly altered from that
which occurs in the uncoupled WC model. Notably, the grains are
moving with the ions and field throughout most of the precursor
region in the IC models, and the width of the precursor is substantially
smaller than in the WC model. From this, we suggest that the criterion 
$\Hallg = 1$ can reasonably be used to separate grains with a continuous
distribution of sizes into two distinct sub-populations of coupled
($\Hallg \simgt 1$) and uncoupled ($\Hallg < 1$) charged grain fluids.

The demarcation between these two sub-populations occurs at the maximum
radius of flux-freezing in the grains, $\agfr$, which is defined as 
the radius $\ag$ for which $\Hallg = 1$. From equation 
(\ref{defHallgeq}) it follows that
\be
\label{agfreq}
\agfr = 1.0 \times 10^{-5}\left(B\over 50~\mu{\rm G}\right)^{1/2}
\left(2 \times 10^{4}~\cc\over \nn \right)^{1/2}
\left(10{\rm K}\over T \right)^{1/2}~{\cm}. 
\ee
Grains with $\ag \leq \agfr$ have $\Hallg \geq 1$; to calculate the 
mass fraction of grains contained in this sub-population, we use
the well-known MRN grain distribution function, 
$f(\ag) \propto \ag^{-3.5}$ (Mathis, Rumpl, and Nordsieck 1977).
Using this distribution function, the mass fraction of grains with
$\ag \leq \agfr$ is
\be
\label{rhogfraceq}
\frac{\rhog(\ag \leq \agfr)}{\rhog} = 
\left(\frac{\agfr}{a_{\rm g,max}}\right)^{1/2}
\frac{\left[1 - \left(a_{\rm g,min}/a_{\rm g,fr}\right)^{1/2}\right]}
{\left[1 - \left(a_{\rm g,min}/a_{\rm g,max}\right)^{1/2}\right]} ~,
\ee
where $a_{\rm g,min}$ and $a_{\rm g,max}$ are the minimum and maximum
grain radii of the distribution. For our cloud/core models
we may take $a_{\rm g, min} = 0.01~\mu{\rm m}$, 
$a_{\rm g,max} = 0.3~\mu{\rm m}$, which yields 
$\rhog(\ag \leq \agfr) \simeq 0.48 \rhog$. 
Hence, if we replace $\rhog$ with $\rhog(\ag \leq \agfr)$ to estimate
the grain magnetosound speed (eq. [\ref{vgmsdefeq}]) in interstellar
plasma with a distribution of grain sizes, $\vgms$ is only slightly
greater than that predicted by equation (\ref{vgmsexpresseq}),
$\approx 7~\kms$ --- which is still considerably lower than the values
of $\vcrit$ determined in the previously noted studies of 
``cooling limited" shocks. It is possible, then, that the 
existence and effects of a ``signal-limiting" speed $\vgms$ on the 
formation of shocks and magnetic precursors described in this paper will
not only be qualitatively but also quantitatively correct even 
when there is a spectrum of grain radii. As stated above, we hope to
definitively resolve this particular issue in future work.


\section{Summary}

The principal results of this paper are as follows:

\begin{enumerate}

\item
We calculated the dispersion relations for compressive waves
in a dusty plasma for two models, where the dust is weakly coupled
and strongly coupled to the magnetic field, respectively.
In the ``weakly coupled'' (WC) model, the only propagating modes
in the charged fluids are ion magnetosound waves at very short
wavelengths and neutral magnetosound waves at very long wavelengths.
In the ``strongly coupled'' (SC) model, grain magnetosound waves appear
at intermediate wavelengths.

\item
We simulated the evolution of small-amplitude, Gaussian wave packets
to see how different plasma components respond to disturbances
on various length scales. When the length scales are small enough for
decoupling of the charged and neutral particles to occur, the dynamics
are dramatically different in the WC and SC models. In the WC model 
there is no wave propagation as such. Propagating ion magnetosound waves
are allowed but strongly damped, and so have virtually no affect on the
dynamics. Rapid diffusion of ions and electrons through the neutral gas
and dust occurs, and the disturbance decays without propagating. In the 
SC model, packets of grain magnetosound waves appear, and the 
disturbance propagates (with weak damping) at the grain magnetosound 
speed. Because the ions and electrons are tied to the magnetic field,
the coupling of grains and magnetic field in the SC model suppresses
diffusive ion motions.

\item
We simulated steady, adiabatic shock waves to compare the effects of 
dust in the WC and SC models. The WC model gives shocks that are 
virtually identical to shocks with no dust: steady shocks are C type for
\Alf\ Mach numbers $\Mna<2.76$ and J type with magnetic precursors in 
the ion-electron fluid for $\Mna>2.76$. We say that steady, adiabatic C
shocks are ``cooling limited'' in the WC model because the C-J 
transition occurs when the neutrals become too hot to remain everywhere 
supersonic.

\item
Steady, adiabatic C shocks are also cooling limited in the SC model
if $\vgms>2.76\vnA$. Their structures resemble their WC counterparts, 
with differences attributable to additional momentum transfer caused by 
grain-neutral scattering.

\item
In the SC model steady, adiabatic C shocks are ``signal limited" if
$\vgms<2.76\vnA$. In this case the C-J transition occurs
when the shock speed exceeds the grain magnetosound speed, as
predicted by Chernoff (1985). Shocks faster than the grain
magnetosound speed have no magnetic precursor in any fluid.

\item
We simulated time-dependent shocks to see how magnetic precursors
form in the WC and SC models. Consistent with our study of 
small-amplitude disturbances, the dynamics are dramatically different in
the two cases. In the WC case, a magnetic precursor forms via diffusion
of the ion-electron plasma through the dust and neutrals.
In the SC case, the precursor forms via the propagation of
grain magnetosound waves.
\end{enumerate}
Our calculations incorporated some very unrealistic assumptions
in order to isolate the essential physics: Real grains have a spectrum
of sizes and real shocks have chemistry and radiative cooling.
Having understood the physics, our next investigation will
determine realistic values of \vcrit\ by relaxing these
assumptions.

\acknowledgements{This work was supported by the New York Center for
Studies on the Origins of Life (NSCORT) and the Department of Physics,
Applied Physics, and Astronomy at Rensselaer Polytechnic Institute,
under NASA grant NAG 5-7589. Comments from an anonymous referee are
acknowledged with appreciation.}

\appendix
\section{Diffusion Coefficient for Ions in the Weakly Coupled (WC) Model}
The ion-magnetic field diffusion coefficient can be most easily obtained
for this particular model by noting that the diffusion mode has the 
following characteristics (1) it has effective balance between magnetic
forces and ion-neutral collisional drag, and (2) the motions of the 
neutral and grain fluids are negligible compared to that of the ions. 
Moreover, as previously noted, we may also (3) neglect the effects of 
quasiparticles in the WC model. With these simplifying assumptions, the
linearized versions of the plasma force (per unit volume) equation and
magnetic induction equation are then, respectively,
\bea
\label{linionmtmeq}
0 &=& -\frac{B_{0}}{4 \pi} \frac{\partial B}{\partial z}
- \left(\frac{\rhoi}{\tdragin}\right)_{0} \vi \\ 
\label{linmaginducteq}
\frac{\partial B}{\partial t} &=& - B_{0}\frac{\partial \vi}{\partial z} 
\eea
(see eqs. [17] and [9] of Paper I). 
Solving for $\vi$ from equation
(\ref{linionmtmeq}) and substituting that into (\ref{linmaginducteq})
yields the linearized magnetic field diffusion equation,
\be
\frac{\partial B}{\partial t} = 
\left(\frac{B^2}{4 \pi \rhoi} \tdragin \right)_{0}
\frac{\partial^2 B}{\partial z^2} ~.
\ee
Fourier-analyzing this equation corresponds to the replacement
$\partial/\partial t \rightarrow -i \omega$, and 
$\partial/\partial z \rightarrow i k$. Finally, using 
$\tdamp = -1/\imag[\omega]$, $\lambda = 2 \pi/k$, 
$\vims = \viA = \Bxo/\sqrt{4 \pi \rhoio}$, and equation (\ref{lambdaimseqa}) yields
the ion diffusion timescale (\ref{iondifftime}). This can also
be written as 
\be
\taudims = \frac{\lambda^2}{{\cal D}_{\rm{ion}}},
\ee
where ${\cal D}_{\rm{ion}}$ is the ion-magnetic field ambipolar 
diffusion coefficient (\ref{difficoefeq}).



\newpage
\clearpage
\typeout{Table 1}

\tablecaption{Model Parameters}

\tablecolumns{5}
\begin{deluxetable}{llcc}

\tablehead{
\colhead{Parameter} &
\colhead{Symbol}    &
\colhead{WC Model}  &
\colhead{SC Model}  
}

\startdata

Preshock density $\left(\cmMMM\right)$ &
$n_{{\rm n}0}$ &
$2 \times 10^{4}$ &
$2 \times 10^{4}$ \\ \\

Preshock magnetic field $\left(\mu{\rm G}\right)$ &
$B_0$ &
$50$ &
$50$ \\ \\

Preshock temperature (K)\tbnm{a} &
$T_{{\rm n}0}$ &
$10$ &
$10$ \\ \\

Cosmic ray ionization rate\tbnm{b} ~~$\left({\rm s}^{-1}\right)$ &
\cri &
$8.4 \times 10^{-19}$ &
$1.0 \times 10^{-17}$ \\ \\

Dust-to-gas mass ratio & $\left(\rho_{\rm g}/\rho_{\rm n}\right)_0$ &
$10^{-2}$ &
$10^{-2}$ \\ \\

Grain radius ($\mu$m)& $\ag$ &
$0.2$ &
$0.05$ \\ \\

\enddata

\tbnt{a}{Assigned arbitrarily, i.e., not determined
by requiring the preshock gas to be in thermal balance.
Assumed the same for all fluids.}

\tbnt{b}{Values selected so that each model has the same degree
of ionization $\left(\nni/\nn \right)_0 = 3.1 \times 10^{-8}.$}

\end{deluxetable}


\typeout{Table 2}
\clearpage

\tablecaption{Properties of the Preshock Plasma}

\tablecolumns{4}
\begin{deluxetable}{lccc}

\tablehead{
\colhead{Quantity} &
\colhead{Symbol} &
\colhead{ WC Model}  &
\colhead{ SC Model}  
}

\startdata

Ion-neutral drag time (yr) &
\tdragin  &
$1.5\times 10^{-2}$ &
$1.5\times 10^{-2}$ \\ \\

Neutral-ion drag time (yr) &
\tdragni &
$4.5\times 10^4$ &
$4.5\times 10^4$ \\ \\

Neutral-grain drag time (yr) &
\tdragng &
$1.1\times 10^5$ &
$3.0\times 10^4$ \\ \\

Grain-neutral drag time (yr) &
\tdraggn &
$1.1\times 10^3$ &
$3.0\times 10^2$ \\ \\

Inverse grain gyrofrequency (yr) &
$\Omega_{\rm g}^{-1}$ &
$4083$ &
$63$ \\ \\

Grain Hall parameter &
\Hallg &
0.26 &
4.2  \\ \\

\gminus\ abundance\tbnm{a}&
\xgm &
$3.7 \times 10^{-13}$ &
$2.2 \times 10^{-11}$ \\ \\

\gzero\ abundance &
\xgz &
$2.2 \times 10^{-14}$ &
$2.7 \times 10^{-12}$ \\ \\

\gplus\ abundance &
\xgp &
$8.0 \times 10^{-17}$ &
$4.8 \times 10^{-15}$ \\ \\

Inverse quasiparticle frequency (yr) &
$\Omega_{\rm qp}^{-1}$ &
$1.4 \times 10^{-1}$   &
$2.3 \times 10^{-3}$   \\ \\

Electrostatic wavelength (cm) &
$\lambdael$ &
$2.4 \times 10^{15}$ &
$4.0 \times 10^{13}$ \\ \\

\enddata

\tbnt{a}{Fractional abundance relative to neutral particles.}

\end{deluxetable}


\typeout{Table 3}
\clearpage

\begin{deluxetable}{lccc}
\tablecaption{Characteristic Speeds}

\tablecolumns{3}

\tablehead{
\colhead{Quantity}       &
\colhead{Symbol}         &
\colhead{\hspace{2em}Value (\kms)}   
}

\startdata

Ion magnetosound speed           &
\vims                            &
860                                \\ \\

Grain magnetosound speed         &
\vgms                            &
5.0                                \\ \\

Neutral magnetosound speed       &
\vnms                            &
0.56                               \\ \\

Neutral \Alf\ speed              &
\vnA                             &
0.50                               \\ \\

Neutral adiabatic sound speed    &
\vnas                            &
0.25                               \\ \\

Neutral isothermal sound speed   &
\vnis                            &
0.19                               \\ \\

\enddata

\end{deluxetable}


\clearpage

\begin{figure}
\figurenum{1}
\plotone{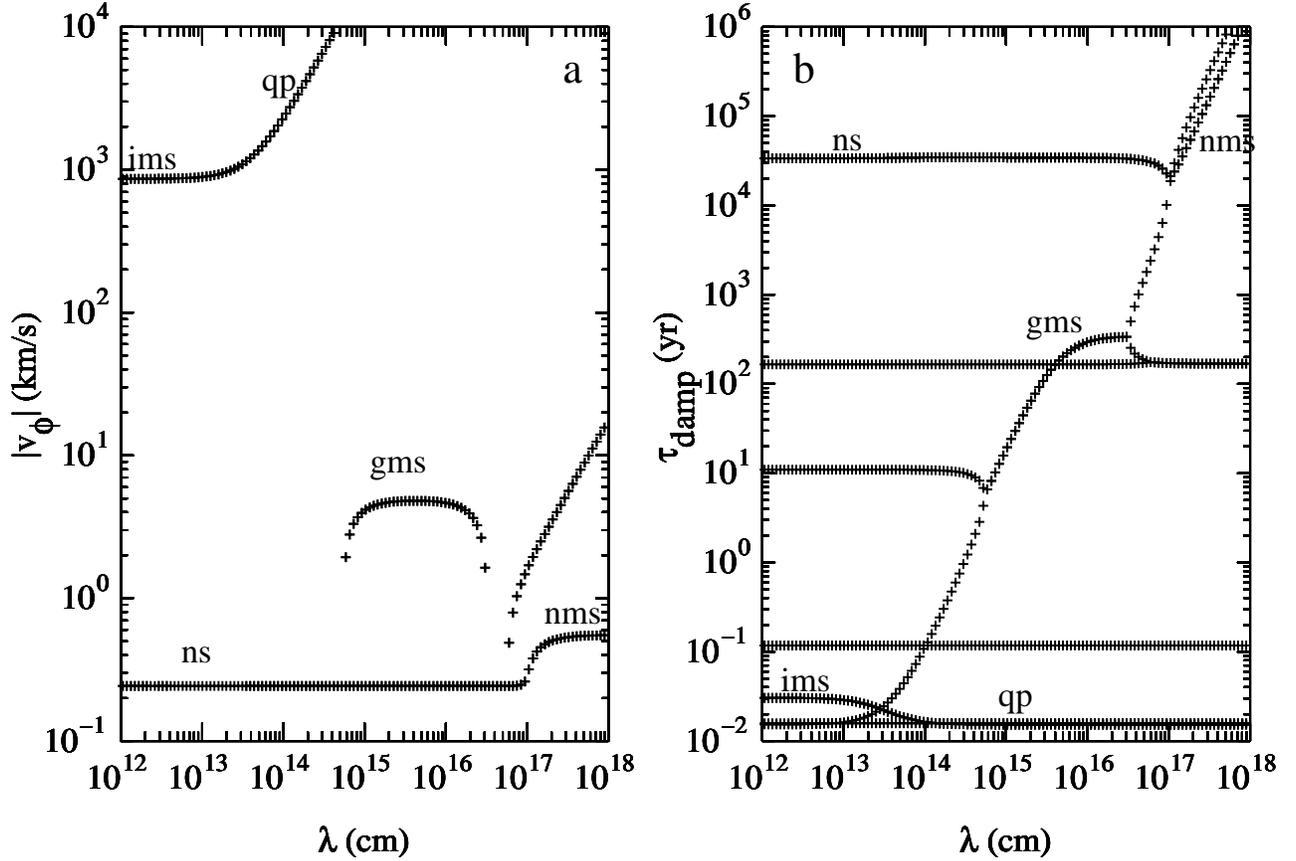}
\caption{Dispersion relation for the WC dust model, where
the grains are weakly coupled to the magnetic field.
({\it a}) Phase velocity $\vphi = \omegar/k$ vs.\ wavelength,
$\lambda$, where $\omegar \equiv \real[\omega]$ and $k = 2 \pi/\lambda$.
Various wave modes are labeled.
Propagating modes include ion magnetosound (ims),
neutral sound (ns), grain magnetosound (gms), and neutral
magnetosound (nms) waves.
The quasiparticle (qp) mode is a nonpropagating oscillation (see text).
({\it b}) Damping time $\tau_{\rm{damp}} = - 1 /\omegai$,
where $\omegai \equiv \imag[\omega]$.
Modes with diffusive damping have a slope
$d(\log \tdamp)/d(\log \lambda) = 2$.}
\end{figure}


\clearpage

\begin{figure}
\figurenum{2}
\plotone{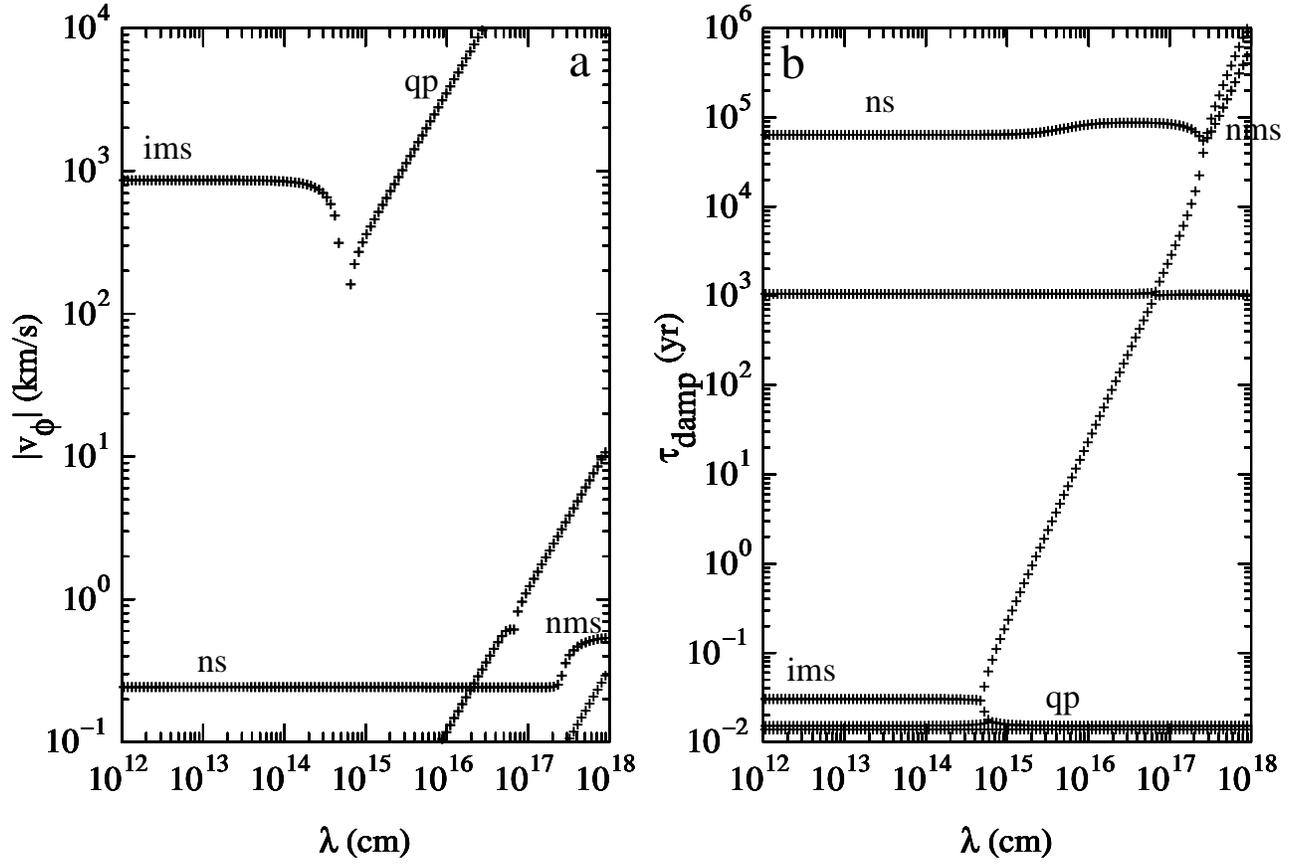}
\caption{Similar to Fig.~1 but for the SC dust model, where
the grains are strongly coupled to the magnetic field.}
\end{figure}


\clearpage

\begin{figure}
\figurenum{3}
\epsscale{0.80}
\plotone{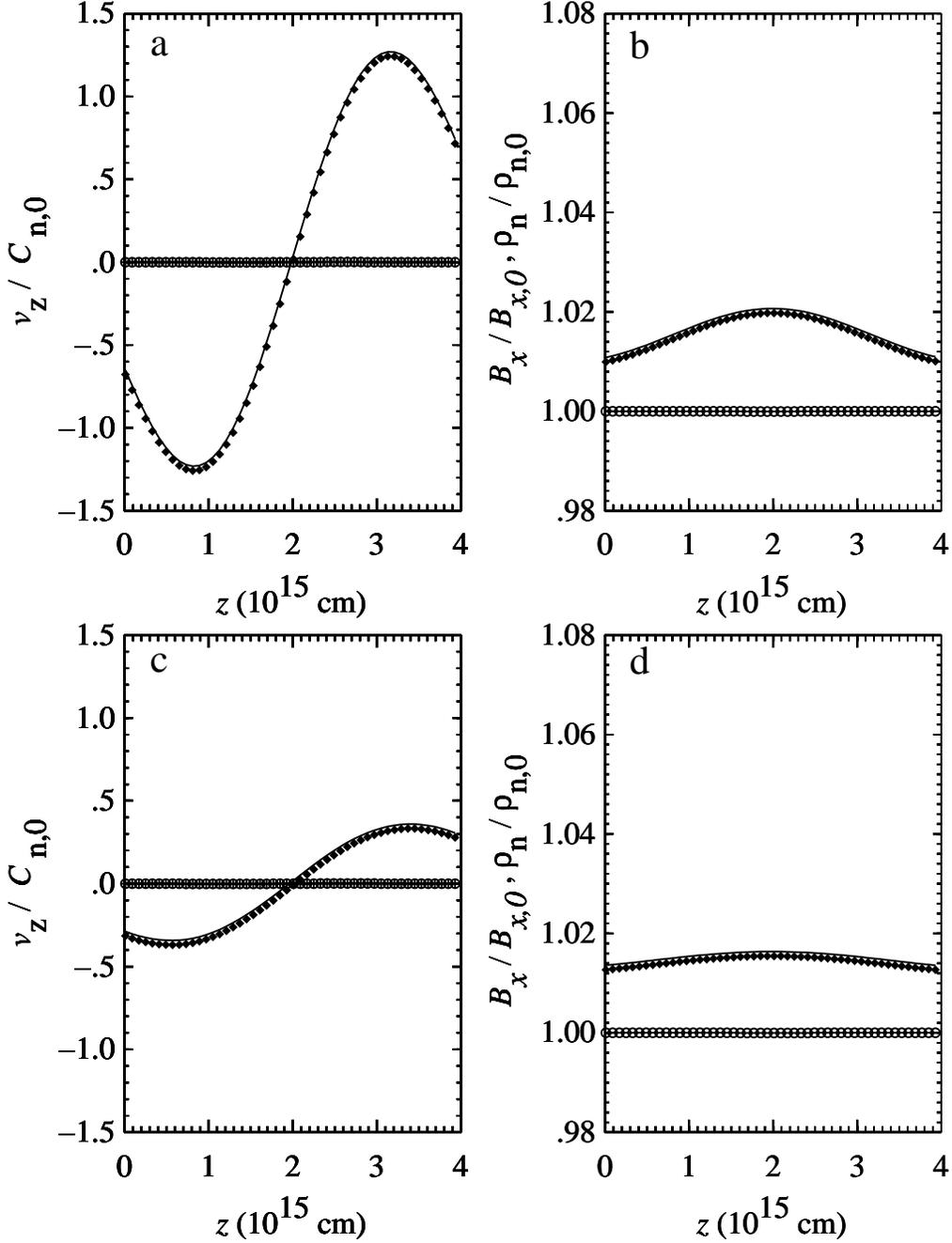}
\caption{Temporal evolution of a Gaussian wavepacket in the WC model.
The initial packet width and amplitude are $\ell_{B} =4 \times 10^{14}~\cm$ and 
$\delta B=0.1B_0$, respectively.
For visual clarity, only every fourth data point is plotted.
({\it a}) Velocities of the ions ({\it filled diamonds}), dust grains ({\it crosses}), and
neutrals ({\it open circles}) at time $t_{1}=8.5~\yr$.
Velocities are in units of the isothermal sound speed
of the undisturbed state ($=0.19~\kms$).
({\it b}) Magnetic field ({\it filled diamonds}) and neutral
density $\rhon$ ({\it open circles}) at time $t_{1}$,
normalized to values for the initial undisturbed state.
({\it c}) Same as ({\it a}) but at a later time $t_{2}=17~\yr$.
({\it d}) Same as ({\it b}), but at time $t_{2}$.}
\end{figure}


\clearpage

\begin{figure}
\figurenum{4}
\epsscale{0.80}
\plotone{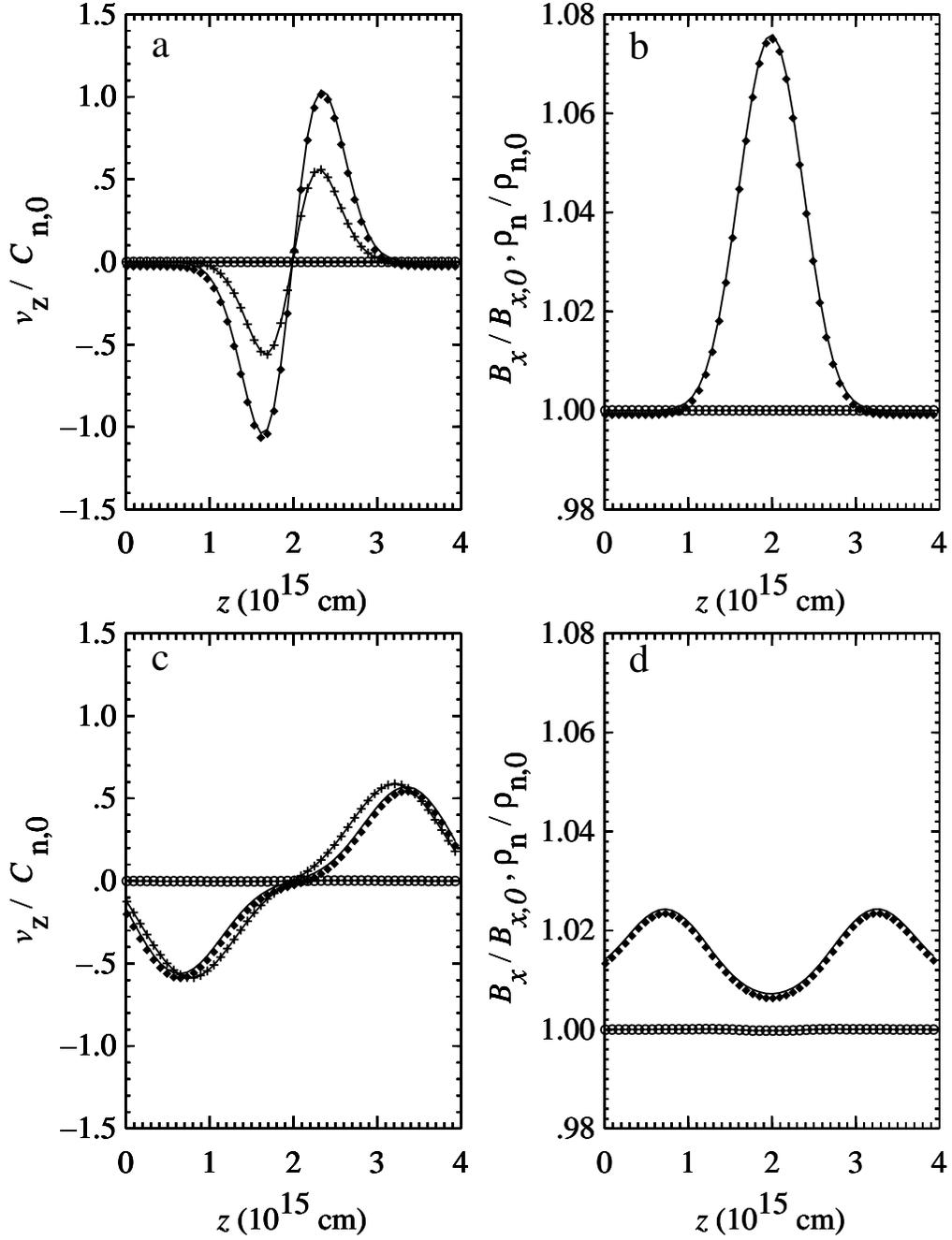}
\caption{Same initial conditions as Fig.~3 but 
computed for the SC model.
({\it a}) Velocities at time $t_{1} = 8.5~\yr$.
({\it b}) Magnetic field and density at time $t_{1}$.
({\it c}) Same as ({\it a}) but at time $t_{2} = 77~\yr$.
({\it d}) Same as ({\it b}) but at time $t_{2}$.}
\end{figure}


\clearpage

\begin{figure}
\figurenum{5}
\epsscale{0.80}
\plotone{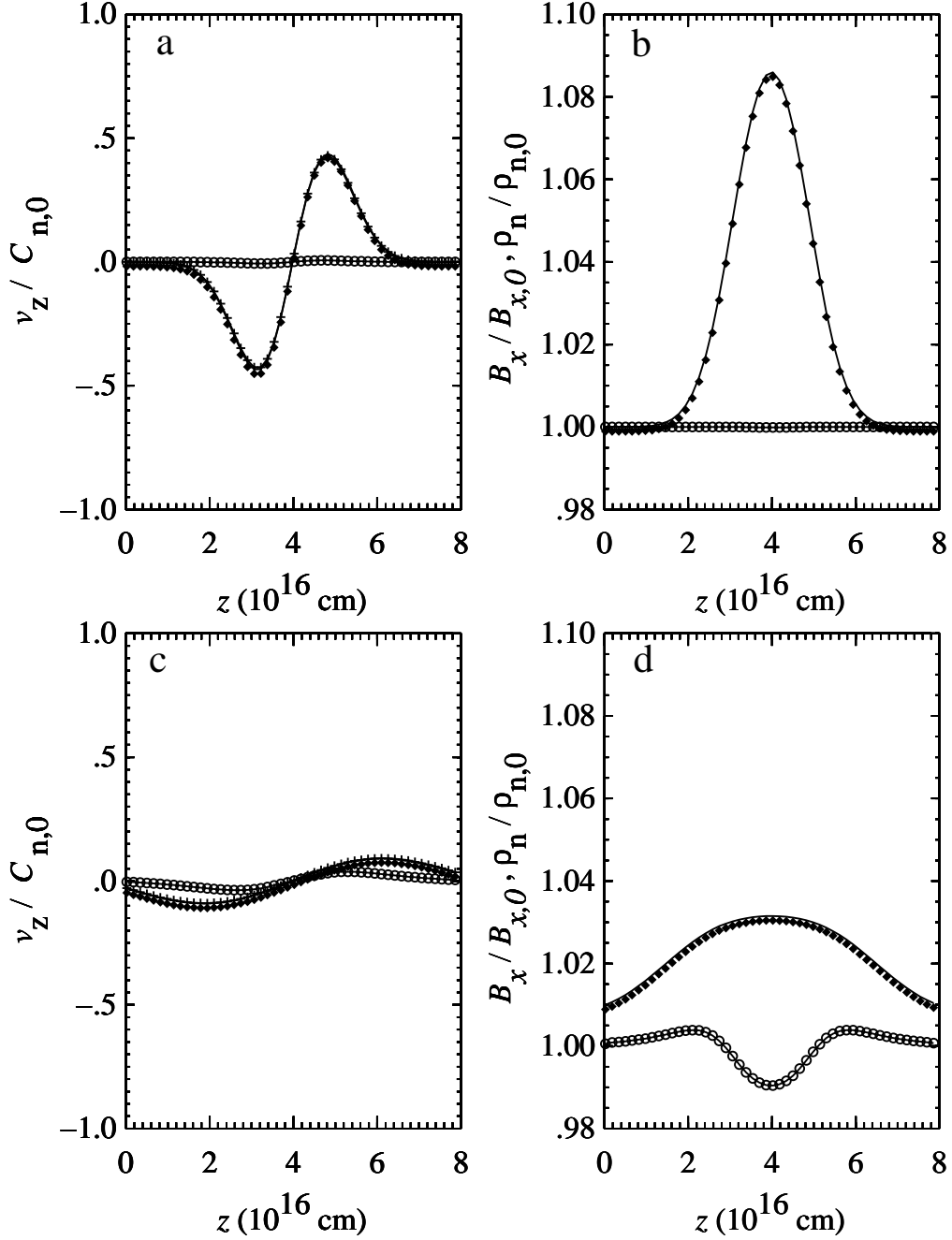}
\caption{Same as Fig.~4 but for a larger initial
packet width: $\ell_B = 1.1 \times 10^{16}~\cm$. ({\it a}) and
({\it b}) are at $t_{1} = 340$ yr, while ({\it c}) and ({\it d}) are
at $t_{2} = 4410$ yr.}
\end{figure}


\clearpage

\begin{figure}
\figurenum{6}
\epsscale{0.80}
\plotone{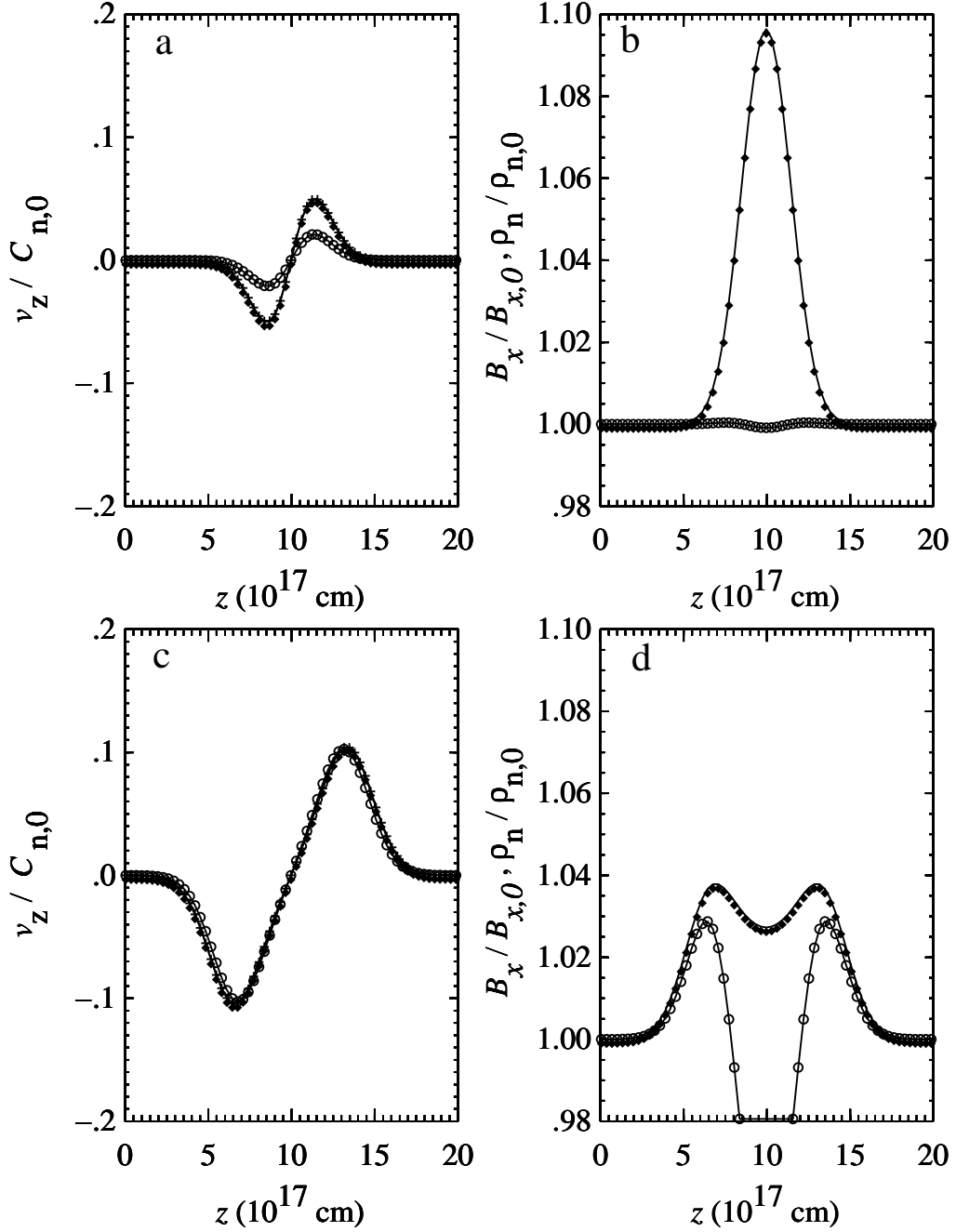}
\caption{Same as Fig.~5 but for an even larger width:
$\ell_B = 2 \times 10^{17}~\cm$. ({\it a}) and ({\it b}) are
at $t_{1}= 1.2 \times 10^4$ yr, while ({\it c}) and ({\it d}) are at
$t_{2} = 1.7 \times 10^5$ yr.}
\end{figure}


\clearpage

\begin{figure}
\figurenum{7}
\epsscale{0.40}
\plotone{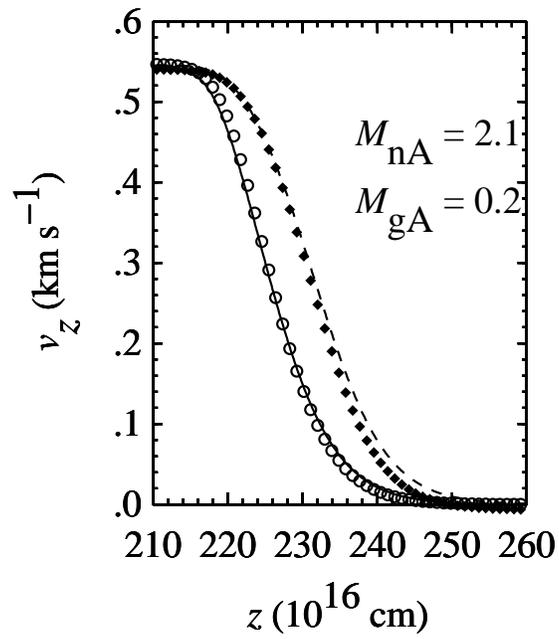}
\caption{Symbols: Numerical predictions of the ion velocity ({\it filled
diamonds}) and neutral velocity ({\it open circles}) in a steady shock 
computed for the WC model. Smooth curves: exact solutions for a steady
shock with the same speed. The neutral and grain \Alf\ Mach numbers are
indicated.}
\end{figure}


\clearpage

\begin{figure}
\figurenum{8}
\epsscale{0.80}
\plotone{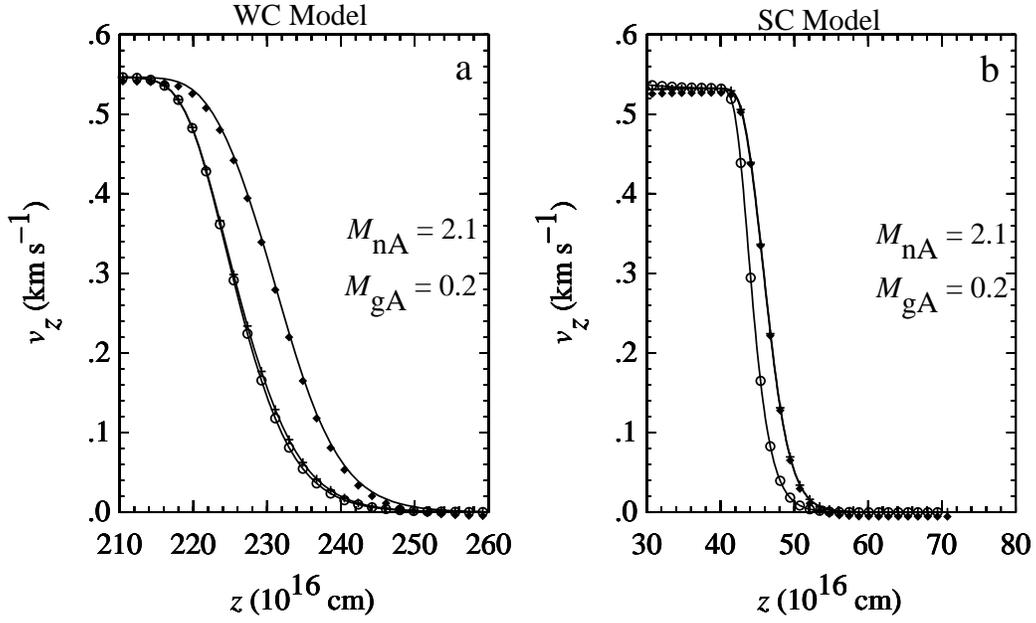}
\caption{Comparison of the velocity profiles for steady, adiabatic shocks
with $\Mna=2.1$ for the WC (Fig.~$8a$) and SC (Fig.~$8b$) models.
The grain \Alf\ Mach number is also indicated. Both solutions are C
shocks.}
\end{figure}


\clearpage

\begin{figure}
\figurenum{9}
\epsscale{0.80}
\plotone{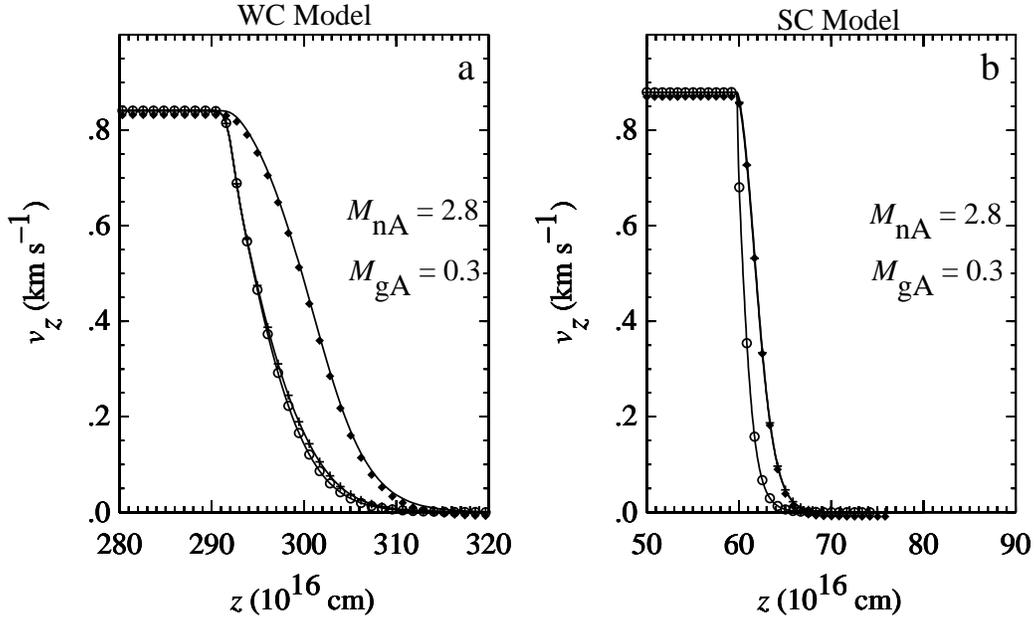}
\caption{Same as Fig.~8 but for $\Mna=2.8$. Both solutions are
weak J shocks. The C-J transition has occured because the
neutral fluid is too hot to remain subsonic.
}
\end{figure}


\clearpage

\begin{figure}
\figurenum{10}
\epsscale{0.80}
\plotone{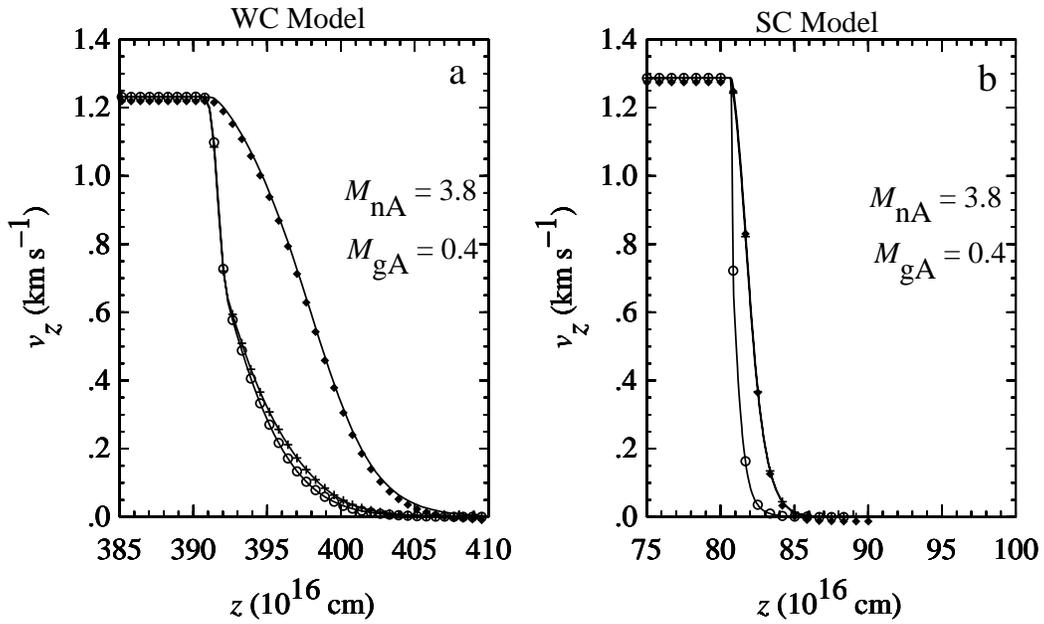}
\caption{Same as Fig.~9 but for $\Mna=3.8$. In both cases,
the solutions are J shocks with magnetic precursors in the charged 
fluid. In the SC model, the grains move with the ions and electrons.
In the WC model, the grains are nearly at rest in the neutral
fluid, with a small velocity difference caused by betatron
acceleration in the magnetic precursor.}
\end{figure}


\clearpage

\begin{figure}
\figurenum{11}
\epsscale{0.80}
\plotone{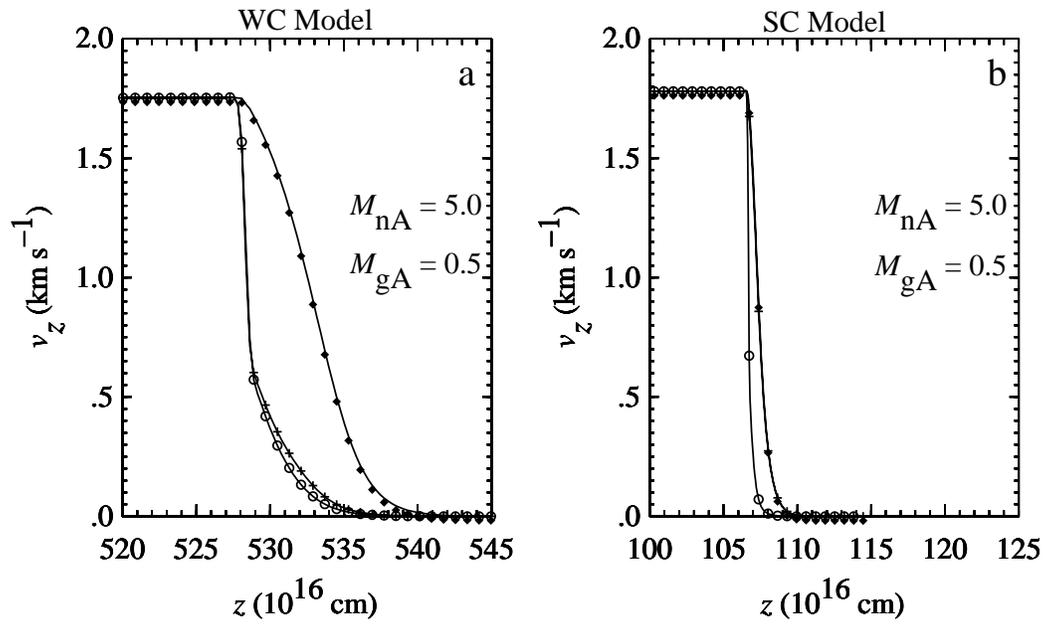}
\caption{Same as Fig.~8 but for $\Mna=5.0$.}
\end{figure}


\clearpage

\begin{figure}
\figurenum{12}
\epsscale{0.80}
\plotone{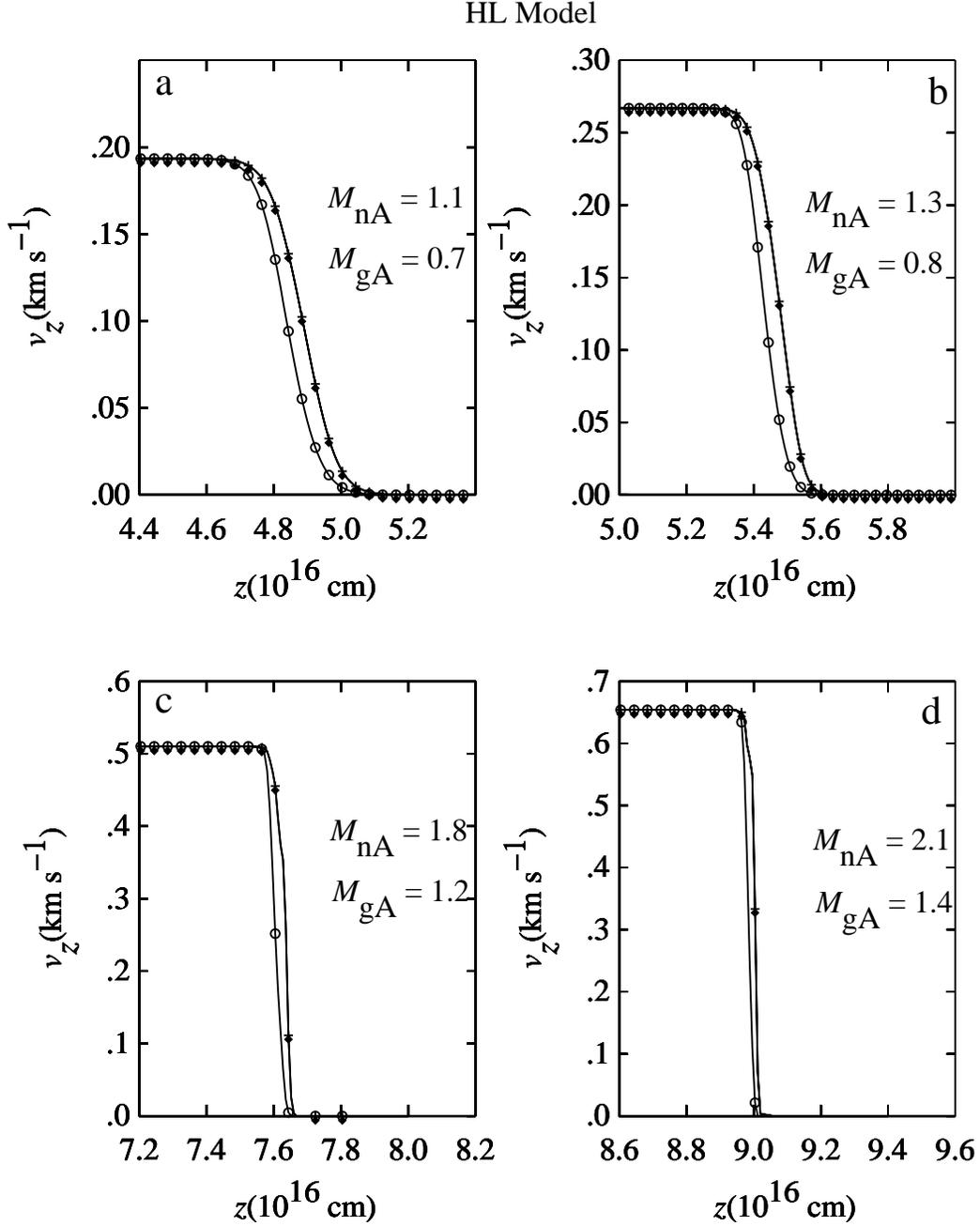}
\caption{Velocity profiles for steady shocks in the ``HL'' model,
where the dust is well coupled to the magnetic field and the grain
magnetosound speed is 0.72 \,\kms. In this case the C-J transition
occurs at $\Mga = 1$, where the shock speed exceeds the signal
speed in the charged fluid. Shocks with $\Mga < 1$ (Fig.~$12a,b$)
are C shocks with magnetic precursors in the charged fluid.
Shocks with $\Mga > 1$ (Fig.~$12c,d$) are J shocks with no magnetic
precursor in any fluid.
(The slight velocity offset between the charged
and neutral velocities in Fig.~$12c,d$ is a numerical artifact.
See text.)}
\end{figure}


\clearpage

\begin{figure}
\figurenum{13}
\epsscale{0.80}
\plotone{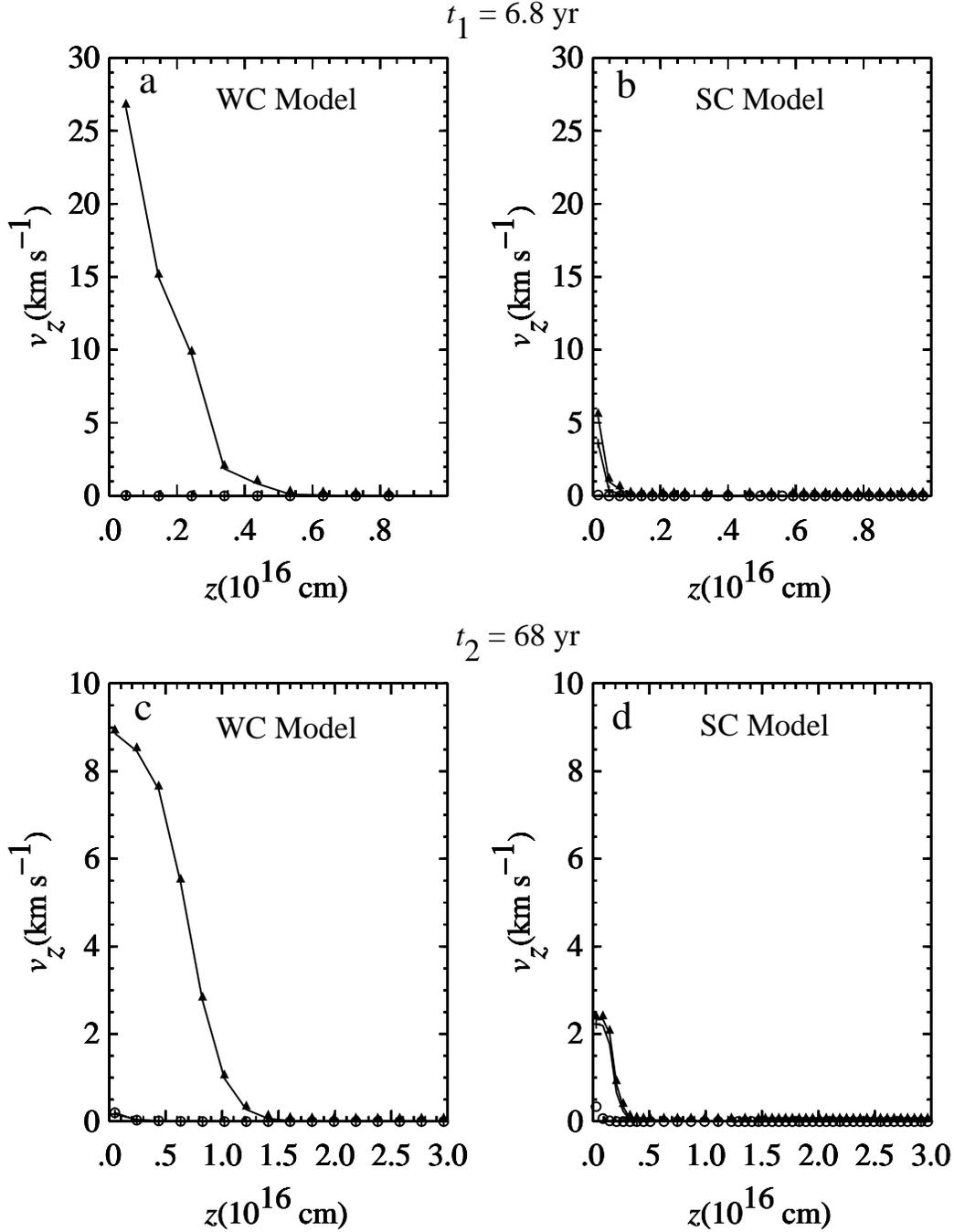}
\caption{Time evolution of a shock in the WC and SC models. At time
$t=0$ the shock discontinuity is located at $z=0$ and is propagating
in the $+z$-direction. ({\it a}) Velocity profile in the WC model at 
$t=6.8$ yr. ({\it b}) SC model at the same time. ({\it c}) WC model
at $t=68$ yr. ({\it d}) SC model at the same time.}
\end{figure}


\clearpage

\begin{figure}
\figurenum{14}
\epsscale{0.80}
\plotone{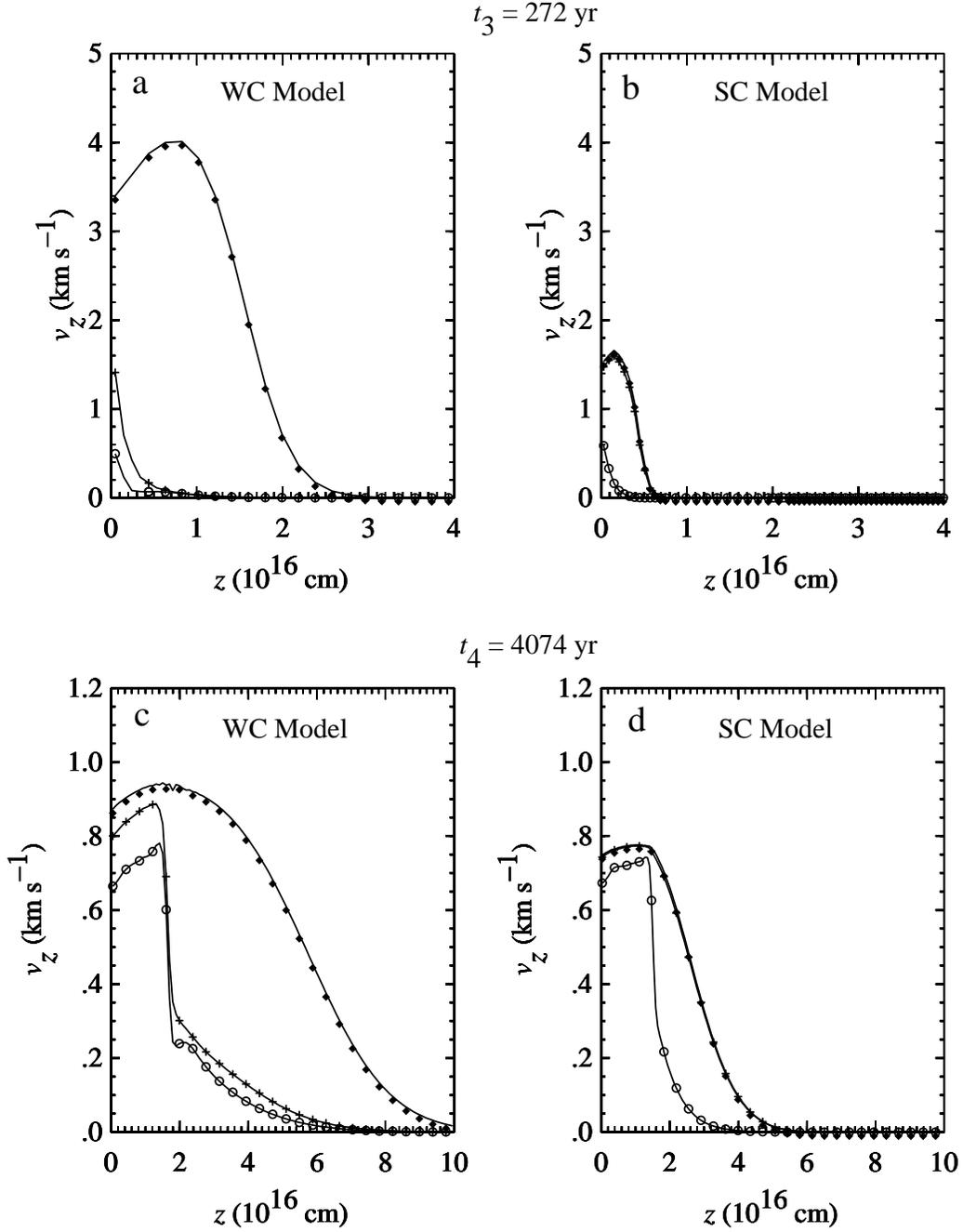}
\caption{Same as in Fig. 13, but at later times. ({\it a}) WC model
at 272 yr. ({\it b}) SC model at the same time. ({\it c}) WC model
at 4074 yr. ({\it d}) SC model at the same time.}
\end{figure}


\clearpage

\begin{figure}
\figurenum{15}
\epsscale{0.80}
\plotone{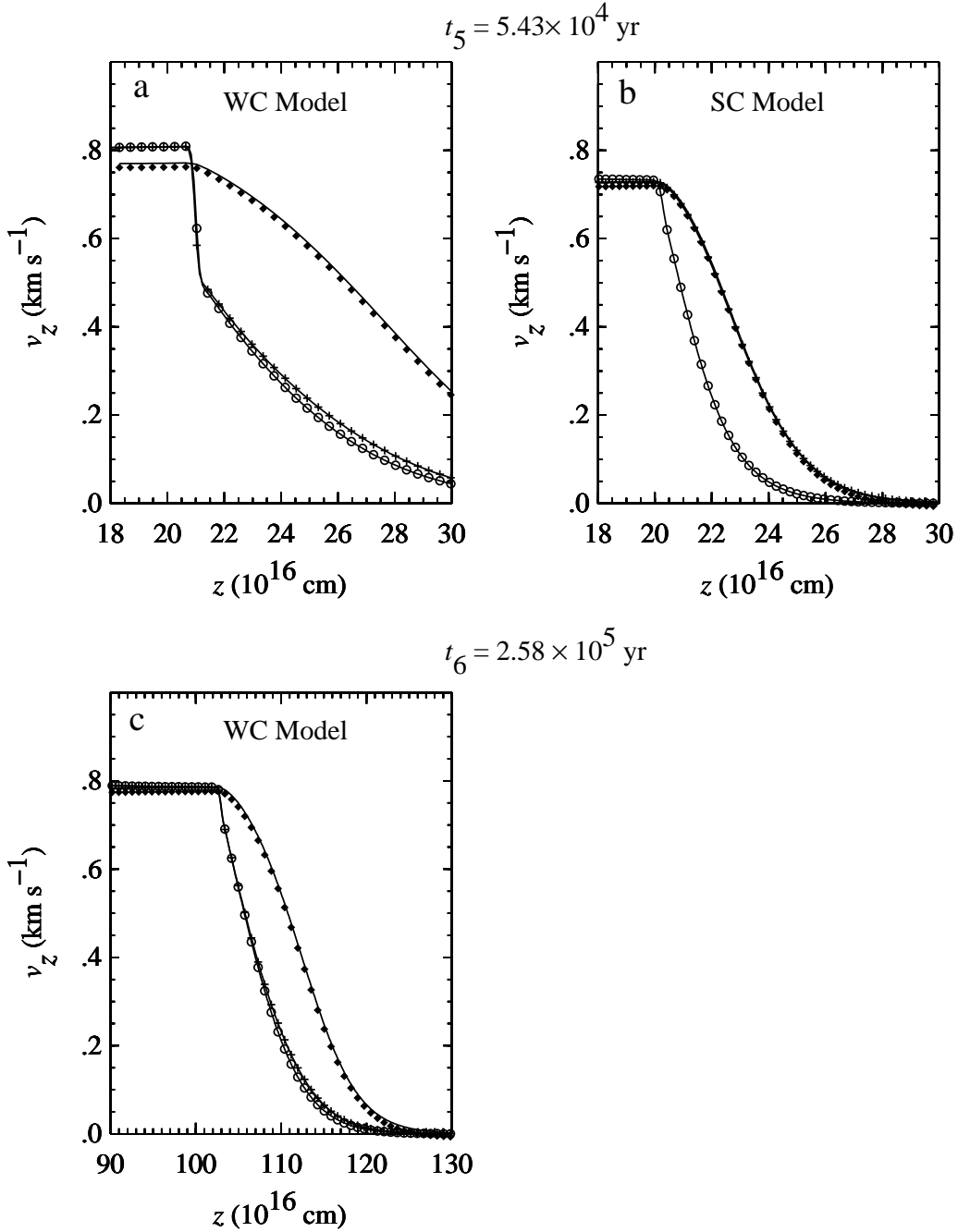}
\caption{Same as in Figs. 13 and 14, but at later times. ({\it a})
WC model at $t=5.43 \times 10^4$ yr. The shock is still in transition,
having the characteristics of a J shock with a magnetic precursor.
({\it b}) SC model at the same time. By this time, the shock has
become a steady C shock. ({\it c}) WC model at $2.58 \times 10^5$ yr.
It is now also a steady C shock by this time. It takes much longer to
form a C shock in the WC model than in the SC model because the 
collisional drag forces (which accelerate the neutrals in the precursor)
in the WC model are less than in the SC model.}
\end{figure}


\clearpage

\begin{figure}
\figurenum{16}
\epsscale{0.40}
\plotone{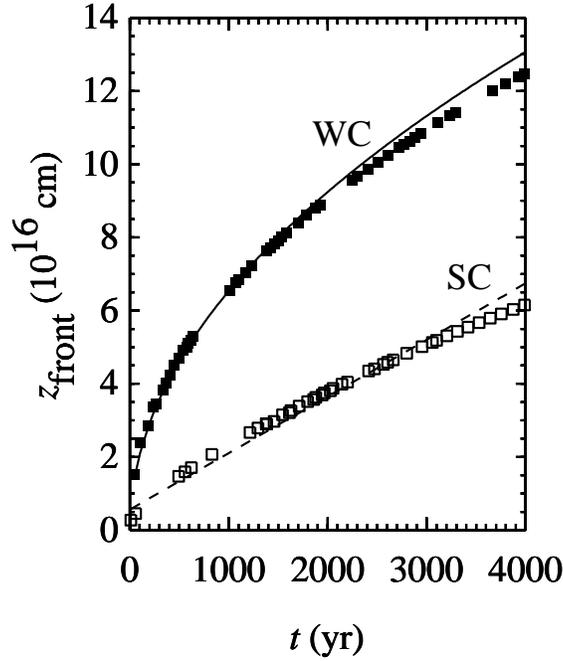}
\caption{Location of the ``front" $z_{\rm front}(t)$ of the 
hydromagnetic signal that propagates from the initial shock 
discontinuity at $z = 0$ and $t = 0$ for the WC ({\it filled boxes}) and
SC ({\it open boxes}) models displayed in Figs. 13 - 15. The location 
of the front is taken to be the point at which the magnetic field 
strength $B$ first increases perceptibly ($\approx 0.5\%$)
above its preshock (i.e., undisturbed) value. The theoretical prediction
of $z_{\rm front} (t)$ for the ion-diffusion mode ($\propto t^{1/2}$, 
see eq. [\ref{zfronteq}]) is displayed as the {\it solid} line. The
{\it dashed} curve corresponds to
$z_{\rm front} = z_{\rm{disp}} + \vgms t$, which is the prediction
for the grain magnetosound mode, displaced by the amount $z_{\rm{disp}}$
to coincide with the location of $z_{\rm front}$ at the time
$t \approx \Omegag^{-1}$.}
\end{figure}

\clearpage

\begin{figure}
\figurenum{17}
\epsscale{0.80}
\plotone{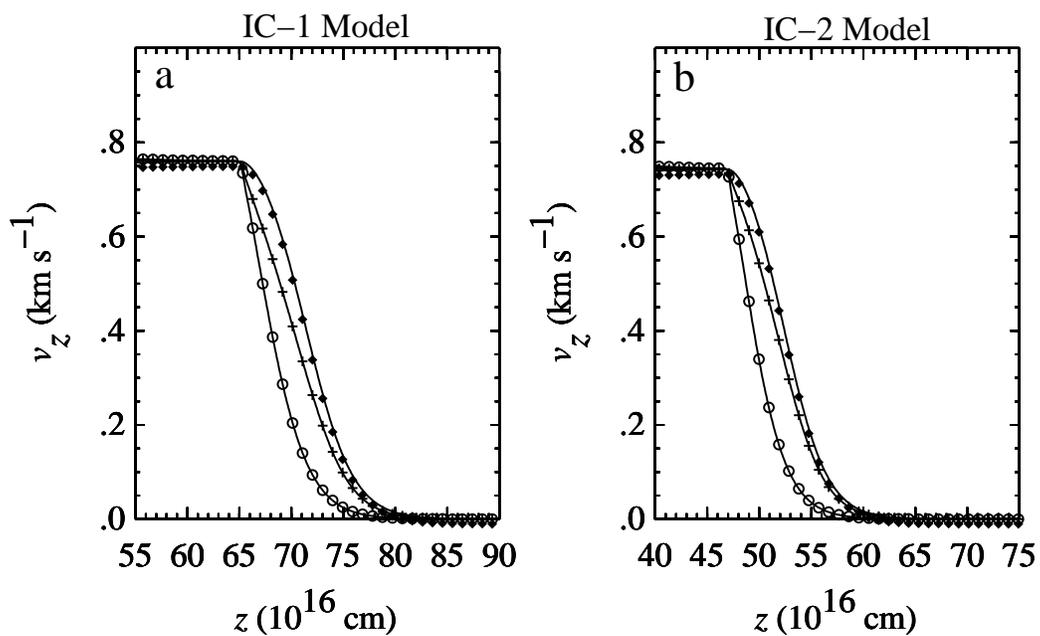}
\caption{C shock flow in ``intermediately-coupled" models. The initial
shock conditions and physical parameters are the same as for the
WC and SC models shown in Figs. 13 - 15, except that the IC-1 model
has grains with $\ag = 0.125~\mu{\rm m}$ ($\Hallg = 1.1$), while
IC-2 has $\ag = 0.09~\mu{\rm m}$ ($\Hallg = 1.3$). ({\it a}) IC-1
model when it has a steady C shock at $t=1.70 \times 10^{5}$ yr.
({\it b}) Steady C shock flow in the IC-2 model at 
$t=1.25 \times 10^{5}$ yr.}
\end{figure}

\clearpage


\end{document}